\documentclass[aps,prapplied,twocolumn,footinbib,longbibliography]{revtex4-2}
\usepackage{graphicx}
\usepackage[utf8]{inputenc}
\usepackage{indentfirst}
\usepackage{physics}
\usepackage{braket}
\usepackage{float}
\usepackage{amsmath}
\usepackage{epstopdf}
\usepackage{footnote} 
\usepackage{CJK}
\usepackage{esint}
\usepackage{color}
\usepackage[T1]{fontenc}
\usepackage{subfigure}
\usepackage{amsfonts}
\usepackage{footmisc}
\usepackage{scrextend}
\usepackage{multirow}
\usepackage[hyperfootnotes=false]{hyperref}
\usepackage[acronym]{glossaries}
\usepackage[normalem]{ulem}

\usepackage[english]{babel}
\usepackage{url}
\usepackage{bm}
\usepackage{hyperref}
\definecolor{darkblue}{rgb}{0,0,0.5}
\hypersetup{
colorlinks=true,
linkcolor=black,
filecolor=blue,
citecolor=darkblue,  
urlcolor=black,
}

\newcommand{\bx}{\boldsymbol x}

\newcommand{\balpha}{\boldsymbol \alpha}

\newcommand{\bZ}{\boldsymbol Z}

\newcommand*\diff{\mathop{}\!\mathrm{d}}

\newcommand {\bbold}{\ensuremath \bm{\hat{b}}}

\newcommand{\calE}{{\cal E}}

\newcommand{\1}{^{(1)}}

\def\be{\begin{equation}}
\def\ee{\end{equation}}
\def\ba{\begin{eqnarray}}
\def\ea{\end{eqnarray}}
\usepackage{bm}

\usepackage{pict2e,picture,graphicx}

\makeatletter
\DeclareRobustCommand{\Arrow}[1][]{%
\check@mathfonts
\if\relax\detokenize{#1}\relax
\settowidth{\dimen@}{$\m@th\rightarrow$}%
\else
\setlength{\dimen@}{#1}%
\fi
\sbox\z@{\usefont{U}{lasy}{m}{n}\symbol{41}}%
\begin{picture}(\dimen@,\ht\z@)
\roundcap
\put(\dimexpr\dimen@-.7\wd\z@,0){\usebox\z@}
\put(0,\fontdimen22\textfont2){\line(1,0){\dimen@}}
\end{picture}%
}
\makeatother
\newcommand{\veryshortrightarrow}{\hspace{.2mm}\scalebox{.8}{\Arrow[.1cm]}\hspace{.2mm}}

\begin{document}

\title{ Teleportation-based microwave-optical quantum transduction enhanced by squeezing}

\author{Jing Wu$^{1}$}
\author{Linran Fan$^{1}$}
\author{Quntao Zhuang$^{2,3}$}
\email{qzhuang@usc.edu}

\address{
$^1$James C. Wyant College of Optical Sciences, University of Arizona, Tucson, AZ 85721, USA
}
\address{
$^2$Ming Hsieh Department of Electrical and Computer Engineering, University of Southern California, Los
Angeles, California 90089, USA
}
\address{
$^3$Department of Physics and Astronomy, University of Southern California, Los
Angeles, California 90089, USA
}

\begin{abstract}
Quantum transduction is an important building block for quantum networking. Although various platforms have been proposed, the efficiency of the-state-of-the-art systems is still way below the threshold to provide robust quantum information transduction via a direct conversion approach. In [Phys. Rev. Applied {\bf 16}, 064044 (2021)], we propose a transduction paradigm based on continuous-variable quantum teleportation that shows a much higher rate in the low cooperativitiy region. While more recently, [Phys. Rev. Research {\bf 4}, L042013 (2022)] proposes to utilize microwave squeezing to assist direct conversion. In this work, we explore the role of squeezing in a teleportation-based transduction protocol and identify a significant performance boost via evaluating quantum capacity lower and upper bounds. Our analyses include both microwave squeezing and optical squeezing, and provide a systematical benchmark between the teleportation-based approach and direct conversion approach. Although with the help of large squeezing, the difference between the teleportation-based protocol and direct conversion protocol becomes smaller, teleportation-based protocol still provides an overall better performance in the practical cooperativity region. In particular, the teleportation-based approach is more robust against imperfect extraction efficiency, even compared with direct conversion with the optimal squeezing. 

\end{abstract}

\maketitle


    
    

\section{Introduction}

Quantum networks transmit and share quantum information between different nodes to enable distributed quantum information processing and utilize unique quantum phenomena such as entanglement and no-cloning for efficient and secure communication~\cite{Acin2007,kimble2008quantum,biamonte2019complex,wehner2018quantum,kozlowski2019towards,zhang2021entanglement,Zhong2021}. Optical photons are ideal for transmitting quantum information over long distances due to its light speed and low loss. However, due to weak optical nonlinearity, quantum information processing heavily relies on superconducting devices in the microwave frequency. To connect microwave information processors with optical quantum links, quantum transduction is required for inter-converting the quantum state between different frequencies~\cite{lauk2020perspectives,awschalom2021development,han2021microwave}.

Various physical systems have been proposed for quantum transduction~\cite{andrews2014bidirectional,bochmann2013nanomechanical,vainsencher2016bi,balram2016coherent,Tsang2010,Tsang2011,fan2018superconducting,xu2020bidirectional,jiang2020efficient,PhysRevLett.103.043603,PhysRevLett.113.203601,shao2019microwave,fiaschi2021optomechanical,han2020cavity,zhong2020proposal,mirhosseini2020superconducting,forsch2020microwave}. Despite the different systems, traditional schemes all utilize a direct conversion approach of transduction: the Hamiltonian can be directly modeled as a beam-splitter interaction between optical and microwave. State-of-the-art performance of such schemes are far from satisfying in neither efficiency or noise level. Recently, microwave single-mode squeezing is proposed to improve the performance of an electro-optical transduction system~\cite{zhong2022quantum}. On the other hand, the teleportation-based approach generates optical-microwave entanglement from the transduction device and then performs quantum teleportation to transmit quantum information, via the time-bin approach~\cite{zhong2020proposal} or the continuous-variable approach~\cite{wu2021deterministic}. In particular, the continuous-variable teleportation approach has shown performance advantages over the direct conversion scheme, especially in the low cooperativity region of near-term devices, thanks to the tunable efficiency and lowered noise~\cite{wu2021deterministic}. At the same time, experimental generation of optical-microwave entanglement on electro-optical device has recently been demonstrated~\cite{sahu2023entangling}, showing great promise for the teleportation-based transduction.

In this paper, we propose to apply single-mode squeezing to enhance the continuous-variable teleportation-based transduction scheme. Via a unified quantum channel modeling of the transduction process, we evaluate the quantum information rates of both teleportation-based and direct conversion schemes assisted by single-mode squeezing. We show that the teleportation-based approach supports a higher quantum information rate than the direct conversion approach. Moreover, the teleportation approach requires less optimum squeezing and is more robust against device imperfections. In particular, teleportation-based transduction supports nonzero rates in a much larger range of extraction efficiencies without the need of any squeezing, versus the direct conversion approach operating at its optimal squeezing condition.

Below, we introduce the organization of the paper.
We describe the electro-optical transduction system and the interaction Hamiltonians assisted by squeezing in Section~\ref{section:cavity electro-optics}. Then we introduce the direct conversion and teleportation-based transduction schemes and present their quantum channel models in Section~\ref{section:channel model for transduction}. In Section~\ref{section:rate comparision}, we compare the capacity of direct conversion and teleportation approaches and show that the latter has larger rates. In Section~\ref{section:fidelity comparison}, we compare the fidelity performance of the transduction protocols in quantum state transfer.

\section{Cavity electro-optics}
\label{section:cavity electro-optics}

\begin{figure}[t]
    \centering
    \includegraphics[width=0.48 \textwidth]{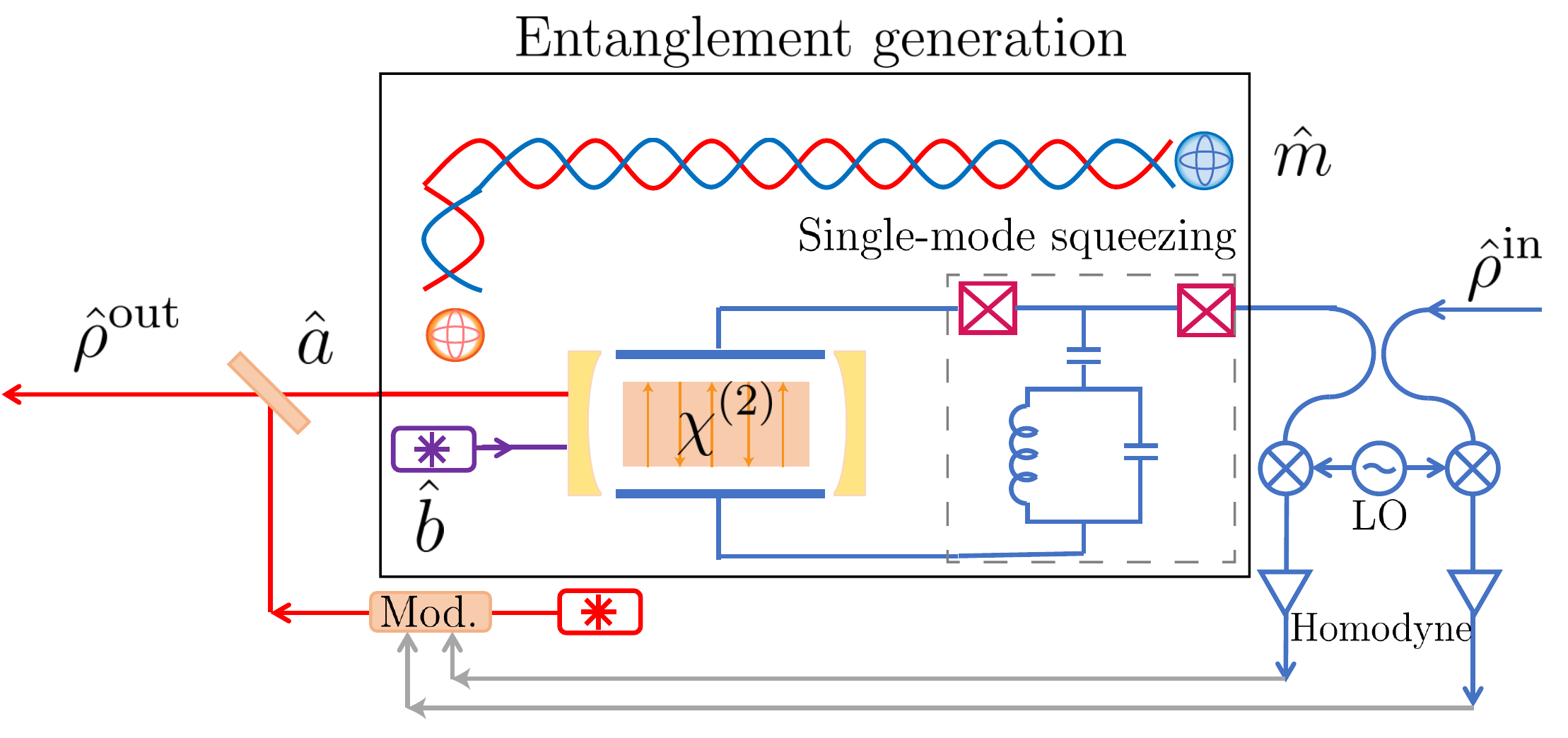}
    \caption{Schematic of the teleportation-based microwave-to-optical transduction scheme. The microwave single-mode squeezing is implemented by the parametric drive of superconducting inductance. The arrows inside the crystal indicate the optical axis of periodic poling structure for optical single-mode squeezing. Density operators $\hat{\rho}^{\rm{in}}$ and $\hat{\rho}^{\rm{out}}$ describe the quantum states of the input and output modes. }
    \label{fig:schematic_cavity}
\end{figure}

Despite that our analyses can be applied to different platforms, we focus on the electro-optical systems due to its state-of-the-art performance~\cite{fan2018superconducting,sahu2022quantum} and simplicity in the interaction Hamiltonian. Also microwave-optical entanglement has recently been generated in such platforms~\cite{sahu2023entangling}, demonstrating an important step towards teleportation-based transduction. As shown in Fig.~\ref{fig:schematic_cavity}, an electro-optical system consists of an optical cavity with $\chi^{(2)}$ nonlinear materials placed between the capacitor of a LC microwave resonator. To enhance the performance, we can apply single-mode squeezing on the microwave side using the inductive nonlinearity in Josephson parametric amplifiers~\cite{castellanos2008amplification,menzel2012path,malnou2018optimal}.
The interaction Hamiltonian of the cavity electro-optics is described by 
\begin{align}
    \hat{H}_{\rm{I}}/\hbar= g_0 (\hat{a}^{\dagger}\hat{b}\hat{m}^{\dagger}+\hat{a}\hat{b}^\dagger\hat{m})+ v(e^{i \theta }\hat{m}^2+e^{-i \theta } \hat{m}^{\dagger 2}),
    \label{eq:Interaction_hamiltonian_MS}
\end{align}
where $\hat{a}$ and $\hat{b}$ are two optical modes and $\hat{m}$ is the microwave mode. Here $g_0$ is the coupling coefficient and $\hbar$ is the reduced Planck's constant. 
The optical (microwave) modes have intrinsic, coupling, and total loss rates $\gamma_{\rm oi}$, $\gamma_{\rm oc}$, and $\gamma_{\rm o}=\gamma_{\rm oi}+\gamma_{\rm oc}$ ($\gamma_{\rm mi}$, $\gamma_{\rm mc}$, and $\gamma_{\rm m}=\gamma_{\rm mi}+\gamma_{\rm mc}$) respectively. The extraction efficiency for the optical (microwave) mode is defined as $\zeta_{\rm o}=\gamma_{\rm oc}/\gamma_{\rm o}$ ($\zeta_{\rm m}=\gamma_{\rm mc}/\gamma_{\rm m}$). To characterize the interaction strength, the interaction cooperativity is defined as 
$
C_g=4 g^2/{\gamma_{\rm o} \gamma_{\rm m}}
$, where we have defined the rescaled coupling coefficient $g=\sqrt{N_{p}}g_0$
with $N_{p}$ being the total intra-cavity pump photon number. Similarly, the squeezing level is defined as $
C_v=4 v^2/{\gamma_{\rm m}^2}
$, where $v$ is the squeezing parameter in Eq.~\eqref{eq:Interaction_hamiltonian_MS}. 

Alternatively, optical squeezing can be realized with parametric down-conversion using the same $\chi^{(2)}$ nonlinear materials between the capacitor~\cite{andersen201630}. As shown in Fig.~\ref{fig:schematic_cavity}, the phase matching condition can be satisfied through geometric dispersion engineering or periodic poling~\cite{boyd2020nonlinear}.
The interaction Hamiltonian is written as
\begin{align}
    \hat{H}_{\rm{I}}/\hbar= g_0(\hat{a}^{\dagger}\hat{b}\hat{m}^{\dagger}+\hat{a}\hat{b}^\dagger\hat{m})+ v(e^{i \theta }\hat{a}^2+e^{-i \theta } \hat{a}^{\dagger 2}),
    \label{eq:Interaction_hamiltonian_OS}
\end{align}
and we can define the corresponding squeezing level $
C_v=4 v^2/{\gamma_{\rm o}^2}
$. 
In both microwave and optical squeezing cases, we assume zero additive noise in the optical frequency
and non-zero additive noise $n_{\rm{in}}$ in the microwave frequency. As explained in our previous work~\cite{wu2021deterministic}, according to the Bose-Einstein distribution, the room-temperature thermal photon at optical frequency is negligible, while the thermal photon $n_{\rm{in}}\sim 10^{-1}$ for a typical $8$ GHZ microwave system cooled to $\sim 10^{-1}$ Kelvin. 

\section{Channel model for transduction}
\label{section:channel model for transduction}
We will consider two types of transduction modality---the traditional direct conversion scheme and the recently proposed continuous-variable teleportation-based scheme~\cite{wu2021}, both enhanced by single-mode squeezing. From the interaction Hamiltonians of Eq.~\eqref{eq:Interaction_hamiltonian_MS} and Eq.~\eqref{eq:Interaction_hamiltonian_OS}, we can solve the quantum channel model of the input-output relation, which turns out to be equivalent to one of the three following single-mode Gaussian channels, thermal-loss (transmissivity $\eta<1$), additive noise (transmissivity $\eta=1$), or thermal amplification channel (gain $\eta>1$). For an input mode $\hat{a}_{\rm in}$, the output mode
\be 
\hat{a}_{\rm out}=
\begin{cases}
\sqrt{\eta}\hat{a}_{\rm in}+\sqrt{1-\eta } \hat{e}, \mbox{ if $\eta<1$},
\\
\hat{a}_{\rm in}+\xi, \mbox{ if $\eta=1$},
\\
\sqrt{\eta}\hat{a}_{\rm in}+\sqrt{\eta-1} \hat{e}^\dagger, \mbox{ if $\eta>1$}.
\end{cases}
\ee 
Here $\xi$ is a complex Gaussian random variable and $\hat{e}$ is a thermal mode with mean photon number $\bar{n}$. We characterize the channel by the transmissivity or gain $\eta$ and the noise mixed into the output mode $N$. In the $\eta\neq 1$ cases, $N=(\bar{n}+1/2)|1-\eta|$, where $1/2$ comes from vacuum fluctuation; At the $\eta=1$ limit, the noise mixed in directly equals the variance of the Gaussian random variable $N$.

\subsection{Direct conversion transduction}

In a direct conversion transduction scheme, the optical mode $\hat{a}$ is coherently pumped with mean photon number $N_p$, which leads to a beamsplitter interaction $ g\hat{b}\hat{m}^\dagger$ in the first part of Eq.~\eqref{eq:Interaction_hamiltonian_MS} or Eq.~\eqref{eq:Interaction_hamiltonian_OS}. Such a beamsplitter interaction enables coherent conversion between the optical mode $\hat{b}$ and the microwave mode $\hat{m}$---achieving the required transduction.
As analyzed in Ref.~\cite{zhong2022quantum},  single-mode squeezing with $C_v>0$ improves the transduction performance in such a direct conversion scheme.

As shown in Appendix~\ref{app:DC_channel}, the transduction channel of direct conversion is equivalent to a single-mode Gaussian channel~\cite{Weedbrook2012}, which can be further simplified to a thermal-loss or thermal-amplification channel. 
In both optical and microwave squeezing cases, the transmissivity or gain of the transduction channel has the identical expression for both microwave-to-optical and optical-to-microwave directions,
\begin{equation}
\eta_{\rm{DC}}=\frac{4C_g \zeta_{\rm{o}}  \zeta_{\rm{m}}}{(1+C_g)^2-4C_v},
\label{eq:transmissivity_DC}
\end{equation}
where the level of squeezing $C_v=4v^2/\gamma_{\rm{m}}^2$ or $C_v=4v^2/\gamma_{\rm{o}}^2$ for the transduction with microwave or optical squeezing.

The noise being mixed in, $N_{\rm{DC}}$, is different for each of the four cases of microwave/optical squeezing and microwave-to-optical/optical-to-microwave directions, and we present the derivations in Appendix~\ref{app:DC_channel_eta_n_}. Overall, $N_{\rm{DC}}$ depends on the cooperativity $C_g$, the level of squeezing $C_v$, the microwave noise $n_{\rm in}$ and the extraction efficiencies $\zeta_{\rm{o}}$ and $\zeta_{\rm{m}}$. We provide the formula for the case of microwave-to-optical (`$\rm m \veryshortrightarrow o$') transduction assisted by microwave squeezing,
\begin{equation}
    N_{\rm{DC,MS}}^{\rm m \veryshortrightarrow o}=\frac{\sqrt{\left[A_-^2+4C_g\zeta_{\rm{o}}B_+\right]\left[A_+^2+4C_g\zeta_{\rm{o}}B_-\right]}}{2\left[(1+C_g)^2-4C_v\right]},
\end{equation}
where
\begin{align}
    &A_{\pm}=1+C_g\pm2\sqrt{C_v},\\
    &B_{\pm}=(1+2n_{\rm{in}})(1-\zeta_{\rm{m}})-1\pm2\sqrt{C_v}.
\end{align}
For the other cases, $N_{\rm{DC}}$ is lengthy and we present their derivations in the appendices.

The transmissivity of Eq.~\eqref{eq:transmissivity_DC} increases as $C_v$ increases at the cost of increased noise. In order for the electro-optical cavity to operate in a stable condition, we require
\begin{equation}
    1+C_g>2\sqrt{C_v},
    \label{eq:stable_condition_DC}
\end{equation}
for both optical squeezing and microwave squeezing (See Appendix~\ref{app:DC_stable_condition}).
Despite the constraint, we note that $\eta_{\rm DC}\ge1$ is possible due to single-mode squeezing, where the equivalent bosonic channel become a thermal amplification channel.

\subsection{Teleporation-based transduction}
In a teleportation-based transduction scheme, the optical mode $\hat{b}$ is coherently pumped with mean photon number $N_{p}$, Eq.~\eqref{eq:Interaction_hamiltonian_MS} and Eq.~\eqref{eq:Interaction_hamiltonian_OS} both lead to a two-mode-squeezing interaction $ g\hat{a}^\dagger \hat{m}^\dagger$ between the optical mode $\hat{a}$ and the microwave mode $\hat{m}$ for entanglement generation. Including the effect from single-mode squeezing, one can solve the quadrature covariance matrix (see definition in Appendix~\ref{app:gaussian_channel}) of the optical-microwave entangled state in a standard form, 
\begin{align}
\mathbf{V}_{\rm{mo}} & = \frac{1}{2}
\begin{pmatrix}
u_q & 0 &v_q & 0   \\
0 & u_p & 0 & -v_p \\
v_q & 0 & w_q & 0\\
0 & -v_p & 0 & w_p
\end{pmatrix}.
\label{V_mo_eq}
\end{align}
For microwave squeezing, we have the parameters (see Appendix~\ref{app:TP_channel})
\begin{align}
    & u_{q/p}= 1+\frac{8\zeta_{\rm{m}}(C_g\pm\sqrt{C_v}+n_{\rm{in}}(1-\zeta_{\rm{m}}))}{(1-C_g\mp2\sqrt{C_v})^2},\\
    & w_{q/p}=1+\frac{8{{C_g}}\zeta_{\rm o}\left[1\mp\sqrt{C_v}+n_{\rm in}\left(1-\zeta_{\rm m}\right)\right]}{(1-C_g\mp2\sqrt{C_v})^2},\\
    & v_{q/p}=\frac{4\sqrt{C_g \zeta_{\rm{m}} \zeta_{\rm{o}}}(1+C_g+2n_{\rm{in}}\left(1-\zeta_{\rm{m}} )\right)}{(1-C_g\mp2\sqrt{C_v})^2}.
    \label{eq:TP_parameters_MS}
\end{align}
For the optical single-mode squeezing case, we can solve the parameters similarly and we present the results in Appendix~\ref{app:TP_covariance_matrix}.

\begin{figure}[t]
    \centering
    \includegraphics[width=0.48 \textwidth]{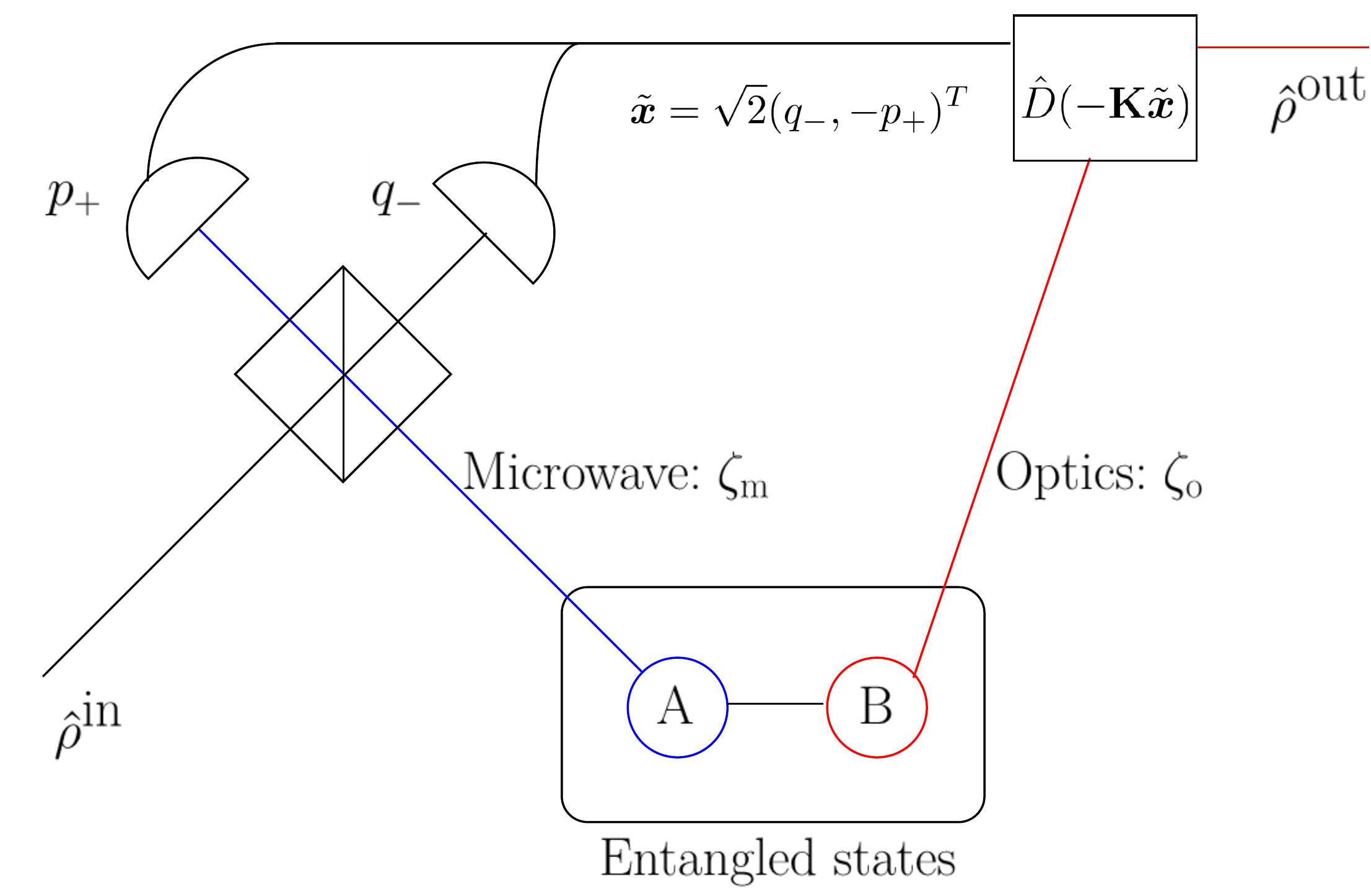}
    \caption{Schematic of continuous-variable quantum teleportation. The displacement is scaled by asymmetric coefficients for different quadratures, $\mathbf{K}=\rm{Diag}(\kappa_q,\kappa_p)$, due to the presence of single-mode squeezing. See main text for more details.}
\label{fig:schematic_teleportation}
\end{figure}

To complete the transduction, one performs continuous-variable quantum teleportation based on the optical-microwave entanglement. Fig.~\ref{fig:schematic_teleportation} shows the case of microwave-to-optical transduction, where the microwave part of the entangled state $A$ is mixed with the input state $\hat{\rho}^{\rm in}$ at microwave frequency by a $50/50$ beamsplitter. Then one performs homodyne detections to obtain results $p_+$ and $q_-$. Based on the results, a scaled displacement of amplitude $\sqrt{2}\left(\kappa_qq_-,\kappa_p p_+\right)$ on q and p quadratures is applied on the optical part to produce the output state $\hat{\rho}^{\rm out}$ at optical frequency. 

As shown in Appendix~\ref{app:TP_channel} , similar to the direct conversion case, the transduction channel of teleportation-based approach can be reduced to a thermal-loss channel 
or a thermal-amplification channel.
Compared to direct conversion, the teleportation channel has a tunable transmissivity (or gain when $\ge1$)
\begin{equation}
    \eta_{\rm{TP}} =\kappa_q \kappa_p,
\end{equation}
where $\kappa_q$ and $\kappa_p$ are the scaling factors of the displacement on $\hat{q}$ and $\hat{p}$ quadrature, respectively. As the entanglement specified by the covariance matrix of Eq.~\eqref{V_mo_eq} is asymmetric between microwave and optical, the channel added noise levels differ for microwave-to-optical (`$\rm m\veryshortrightarrow o$') transduction and optical-to-microwave (`$\rm o\veryshortrightarrow m$') transduction, which are given by
\begin{align}
    &N_{\rm{TP}}^{\rm m\veryshortrightarrow o}=\frac{1}{2}\sqrt{\left(u_q \kappa_q^2-2 v_q \kappa_q +w_q\right)\left(u_p \kappa_p^2-2 v_p \kappa_p +w_p\right)},\\
    &N_{\rm{TP}}^{\rm o\veryshortrightarrow m}=\frac{1}{2}\sqrt{\left(w_q \kappa_q^2-2 v_q \kappa_q +u_q\right)\left(w_p \kappa_p^2-2 v_p \kappa_p +u_p\right)}.
\end{align}
Note that the parameters $u_{q/p}$, $v_{q/p}$ and $w_{q/p}$ have different interpretations for optical squeezing and microwave squeezing.

Similar to the direct conversion case, the teleportation-based approach is subject to a stable condition 
\begin{equation}
    1-C_g>2\sqrt{C_v},
    \label{eq:stable_condition_TP}
\end{equation}
for both optical squeezing and microwave squeezing.
See Appendix~\ref{app:TP_stable_condition} for a derivation. 

\subsection{Choice of squeezing based on direction of transduction}

\begin{figure}[t]
    \centering
    \includegraphics[width=0.45\textwidth]{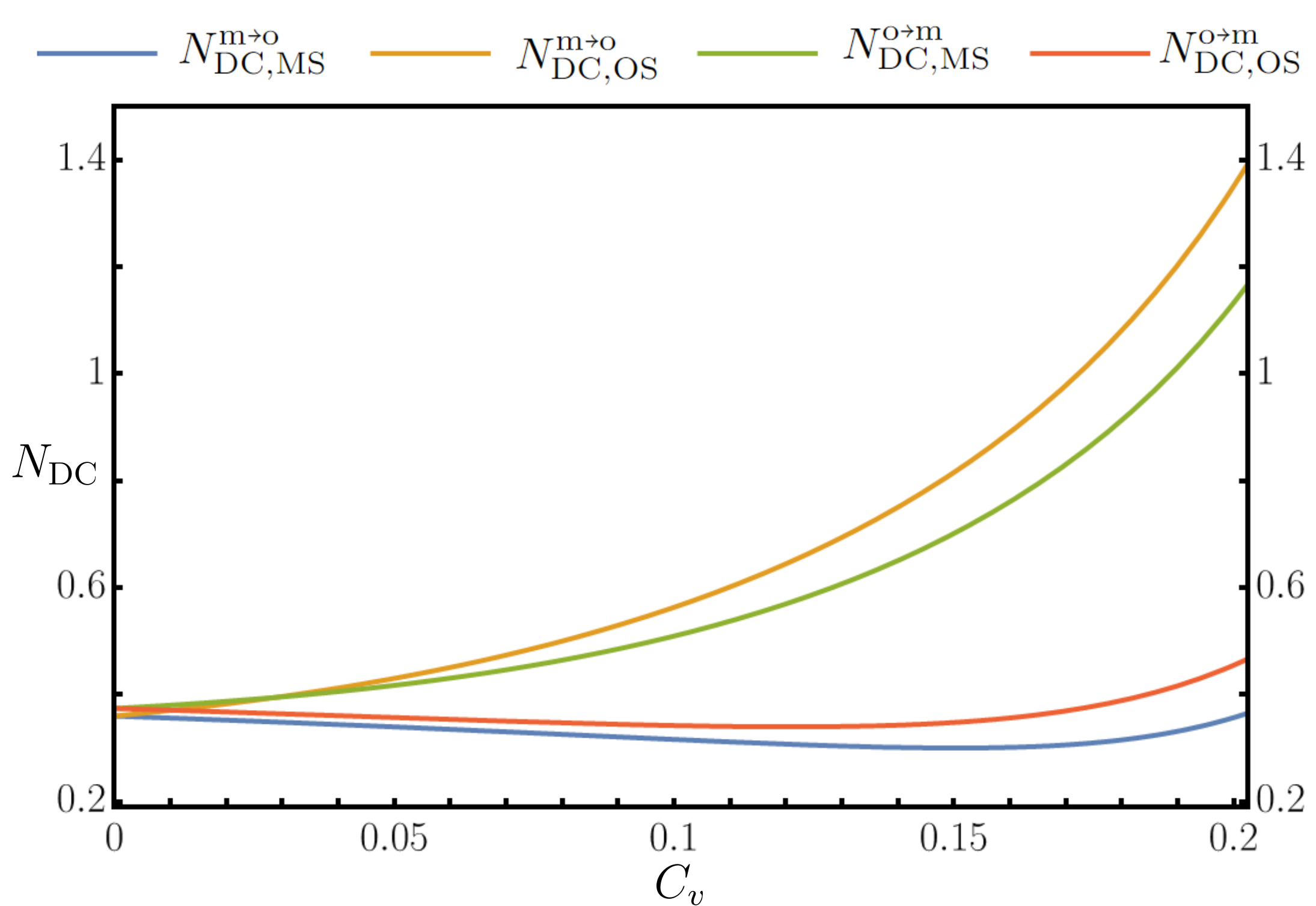}
    \caption{Thermal noise $N_{\rm{DC}}$ versus the squeezing level $C_v$ with microwave thermal photon $n_{\rm{in}}=0.1$, microwave extraction coefficiency $\zeta_{\rm{m}}=0.95$, optical extraction coefficiency $\zeta_{\rm{o}}=0.9$ and cooperativity $C_g=0.1$.}
    \label{fig:Ndc}
\end{figure}

Four different scenarios of transduction arise when one applies either optical or microwave squeezing in addition to the two direction (optical-to-microwave/microwave-to-optical) of transduction. For simplicity, we hope to focus on the operational modes that best suit the direct conversion approach. Therefore, in this section we explore the preferred modality of squeezing for different transduction directions in the direct conversion protocol. 
Although such an optimal choice of squeezing for direct-conversion approach may not be optimal for teleportation-based approach, it suffices to demonstrate teleportation-based approach's advantage over direct conversion approach.

In our comparison, we assume the same set of parameters---the squeezing level $C_v$, the extraction coefficients $\zeta_{\rm{m}}$ and $\zeta_{\rm{o}}$ and the cooperativity $C_g$. Then, the transmissivity with optical and microwave squeezing are the same. For microwave-to-optical transduction, the difference between the noise being mixed in $N_{\rm{DC,OS}}^{\rm m \veryshortrightarrow o}$ and $N_{\rm{TP,MS}}^{\rm m \veryshortrightarrow o}$ is always positive since
\begin{equation}
   {(N_{\rm{DC,OS}}^{\rm m \veryshortrightarrow o})}^2- {(N_{\rm{DC,MS}}^{\rm m \veryshortrightarrow o})}^2=16\frac{(1-C_g^2)C_v(1-\zeta_{\rm{o}})\zeta_{\rm{o}}}{\left((1+C_g)^2-4C_v\right)^2}.
\end{equation}
However, for optical-to-microwave transduction, the noise being mixed in with optical squeezing may not be smaller than that with microwave squeezing. Nevertheless, as shown in Fig.~\ref{fig:Ndc}, in the practical parameter region of optical-to-microwave transduction, the noise being mixed in is smaller with optical squeezing.
To achieve the best performance, from now on, we will consider the microwave squeezing for microwave-to-optical transduction and optical squeezing for optical-to-microwave transduction.

\section{Rate comparison between direct conversion and teleportation}
\label{section:rate comparision}

\begin{figure}[t]
    \centering
    \includegraphics[width=0.47\textwidth]{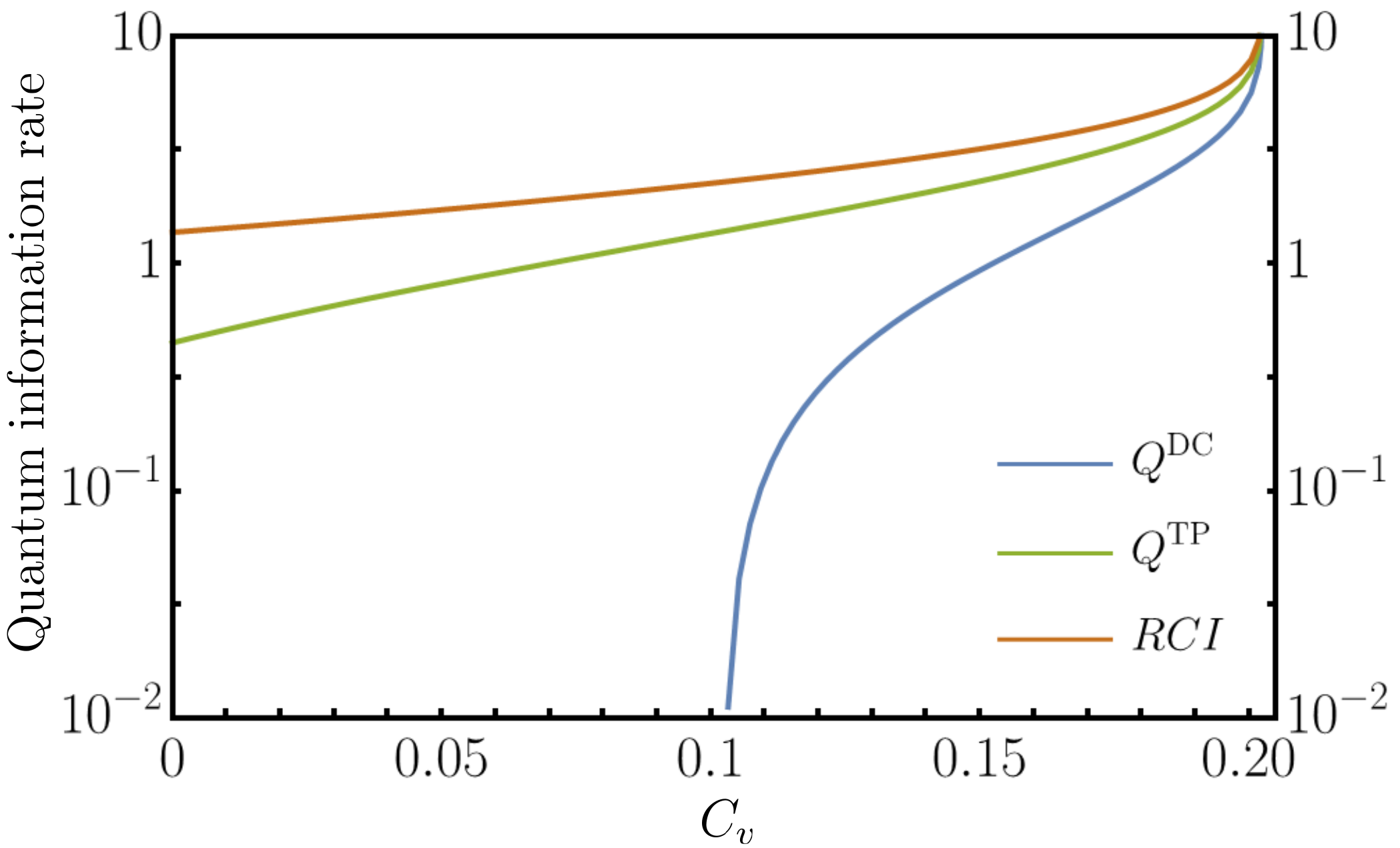}
    \caption{Quantum information rate of direct conversion, teleportation channel and the reverse coherent information (RCI) when $\zeta_{\rm{o}}=\zeta_{\rm{m}}=1$. $Q^{\rm DC}$ is the exact capacity of direct conversion. $Q^{\rm TP}$ is the capacity lower bound of the teleportation-based approach. RCI quantifies the entanglement utilized in the teleportation-based scheme.}
    \label{fig:line_m1o1}
\end{figure}

As we have demonstrated in previous sections, the input-output relation of the direct conversion and teleportation-based transduction schemes can be modeled as thermal-loss or thermal-amplification channels. Ideal transduction requires near unity transmissivity ($\eta\sim1$) and low thermal-noise ($\Bar{n}\sim0$). Compared with direct conversion, the teleportation approach has a tunable transmissivity while possibly having larger noise.
To evaluate the channel's performance, the quantum capacity serves as the metric, representing the highest communication rate when transmitting information through a noisy channel. While the true quantum capacity of a general single-mode Gaussian channel remains unknown, lower bounds and upper bounds of the quantum capacity are available~\cite{holevo2001evaluating,pirandola2017fundamental,fanizza2021estimating}. The equations are summarized in Appendix~\ref{app:gaussian_channel}. These bounds can provide insights into the performance of the transduction schemes. In particular, if the regions formed between lower and upper bounds are well separated for the different schemes, we can definitively determine which approach is superior. This section is divided into two parts: the ideal extraction efficiency case in Section~\ref{subsection:ideal case} and practical extraction efficiency case in Section~\ref{subsection:practical case}. In the practical case, we discuss about cases with low cooperativity and high cooperativity in section~\ref{subsection:practical low cooperativity} and section~\ref{subsection:practical high cooperativity}. 

\begin{figure}[t]
    \centering
    \includegraphics[width=0.48\textwidth]{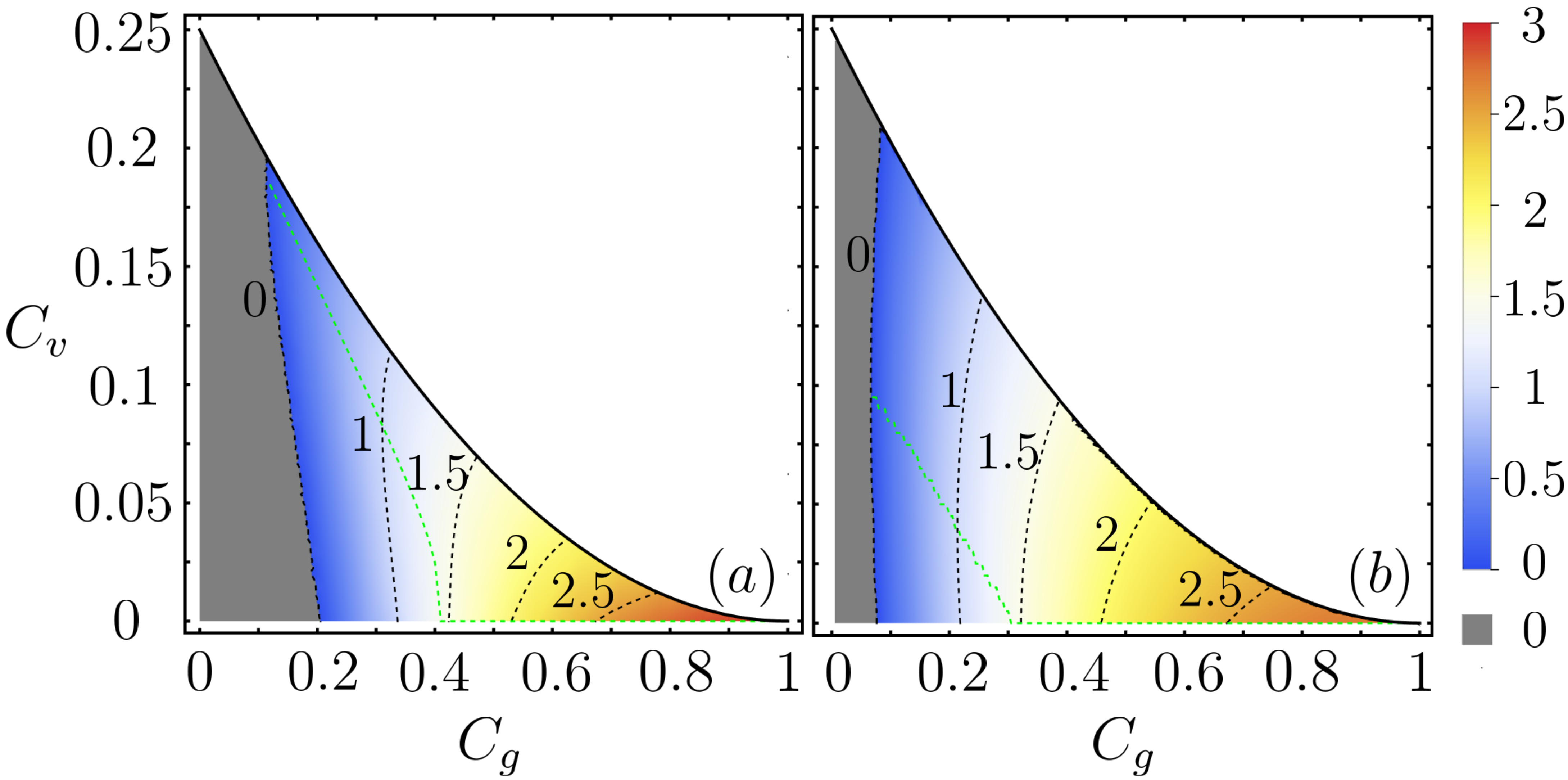}
    \caption{Contour of quantum information rate for microwave-to-optical transduction with $\zeta_{\rm{m}}=\zeta_{\rm{o}}=0.95$ and $n_{\rm{in}}=0.1$. (a) Quantum capacity lower bound of direct conversion transduction. (b) Quantum capacity lower bound of teleportation transduction. The green dashed line corresponds to optimum squeezing $C_v$. 
    \label{fig:contour_MtoO}
    }
\end{figure}

\subsection{Ideal extraction efficiency case}
\label{subsection:ideal case}
We begin the comparison between the channel capacities of the direct conversion and teleportation approaches by examining the ideal scenario with unit extraction efficiencies $\zeta_{\rm{m}}=\zeta_{\rm{o}}=1$.  In this case, the capacity lower bound and upper bound of the direct conversion coincide because, in this case $\Bar{n}_{\rm{DC}}=0$ and the channel becomes a pure loss channel or quantum-limited amplification channel. Furthermore, the capacities of direct conversion in both the optical-to-microwave and microwave-to-optical directions are identical due to the same transmissivity $\eta_{\rm{DC}}$. The capacity of direct conversion, with practical cooperativity $C_g=0.1$, is plotted as the blue line in Fig.~\ref{fig:line_m1o1} against the squeezing parameter $C_v$. For teleportation, we optimize the tuning parameters $\kappa_q$ and $\kappa_p$ to achieve the maximum lower bound of the quantum capacity. The optimization of lower bound of optical-to-microwave transduction and microwave-to-optical transduction coincides. 
Since quantum teleportation utilizes the entanglement of the state, we also include the reverse coherent information (RCI) as the orange curve, which denotes the achievable rate in quantum communication allowing arbitrary local operations on the entanglement described by Eq.~\eqref{V_mo_eq}. As expected, RCI is higher than the lower bound of capacity in the teleportation-based approach. As shown in Fig.~\ref{fig:line_m1o1}, the capacity of the direct conversion is smaller than that of the teleportation at the same level of squeezing $C_v$, which indicates the better performance of the teleportation. Note that the quantum information rate diverges as $C_v$ approaches $(1-C_g)^2/4$. This occurs because, in both direct conversion and teleportation, the transmissivity $\eta \rightarrow 1$ and noise $\bar{n}\rightarrow0$ in this limit and we are evaluating the capacity without energy constraints. 

\begin{figure}[t]
    \centering
\includegraphics[width=0.48\textwidth]{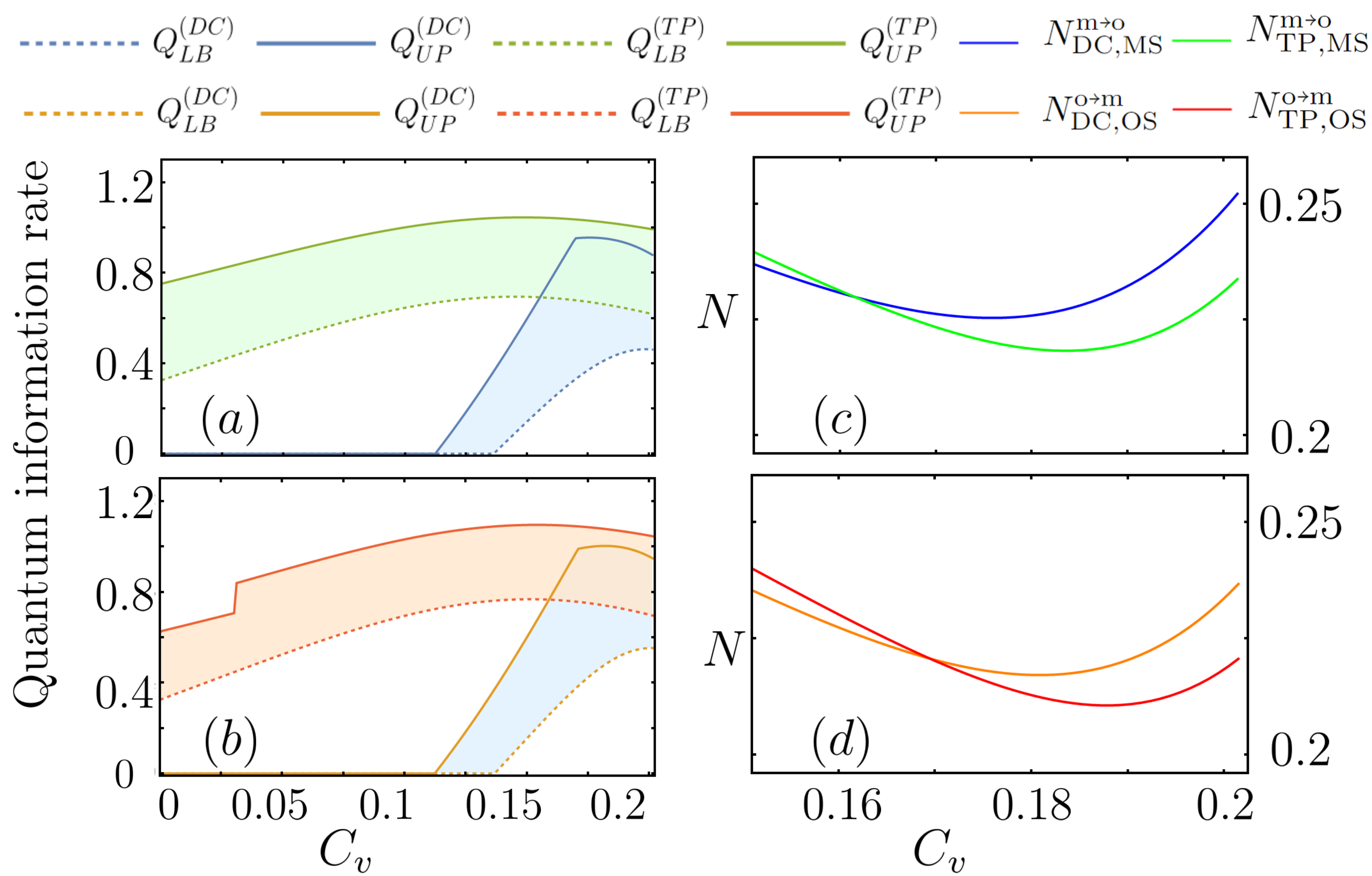}
    \caption{Quantum information rate and noise being mixed in versus squeezing $C_v$ for $\zeta_{\rm{m}}=\zeta_{\rm{o}}=0.975$, $n_{\rm{in}}=0.1$, and $C_g=0.1$. (a) and (c) are microwave-to-optical transduction, (b) and (d) are optical-to-microwave transduction. In the legend, `LB' indicates lower bound and `UB' indicated upper bound. `DC' indicates direct conversion and  `TP' indicates teleportation-based conversion. For the effective noise $N$, `MS' indicates microwave squeezing and `OS' indicates optical squeezing.}
    \label{fig:line_combined}
\end{figure}

\subsection{Practical extraction efficiency case}
\label{subsection:practical case}

We now examine the practical case where the extraction efficiencies $\zeta_{\rm m}$ and $\zeta_{\rm o}$ are below unity.
To fully explore the impact of different levels of cooperativity and squeezing on the quantum information rate of the two transduction approaches, we present a contour plot in Fig.~\ref{fig:contour_MtoO}. The gray area represents zero capacity, indicating no reliable quantum information transmission, while the green dashed line indicates the optimum squeezing $C_v$ that maximizes the lower bound at each fixed cooperativity value $C_g$. We observe that the teleportation approach (in sub-panel (b)) has a larger region of non-zero rate compared to direct conversion (in sub-panel (a)), highlighting its robustness to low cooperativity. When $C_g$ is small, choosing an optimal level of squeezing can enhance the communication rate for both direct conversion and teleportation, as evidenced by the green dashed lines. Notably, the teleportation approach requires a lower level of optimal squeezing compared to direct conversion. As cooperativity increases, the optimum squeezing approaches zero, suggesting that squeezing does not aid in either direct conversion or teleportation. Although we present the results in microwave-to-optical transduction, the rate of optical-to-microwave transduction exhibits similar trends. 

In the following analysis, we focus on two specific regions: the low cooperativity region and the high cooperativity region. For simplicity, we assume equal extraction efficiencies---$\zeta_{\rm{m}}=\zeta_{\rm{o}}=\zeta$---in this section.

\subsubsection{Low cooperativity}
\label{subsection:practical low cooperativity}
\begin{figure}[t]
    \centering
    \includegraphics[width=0.48\textwidth]{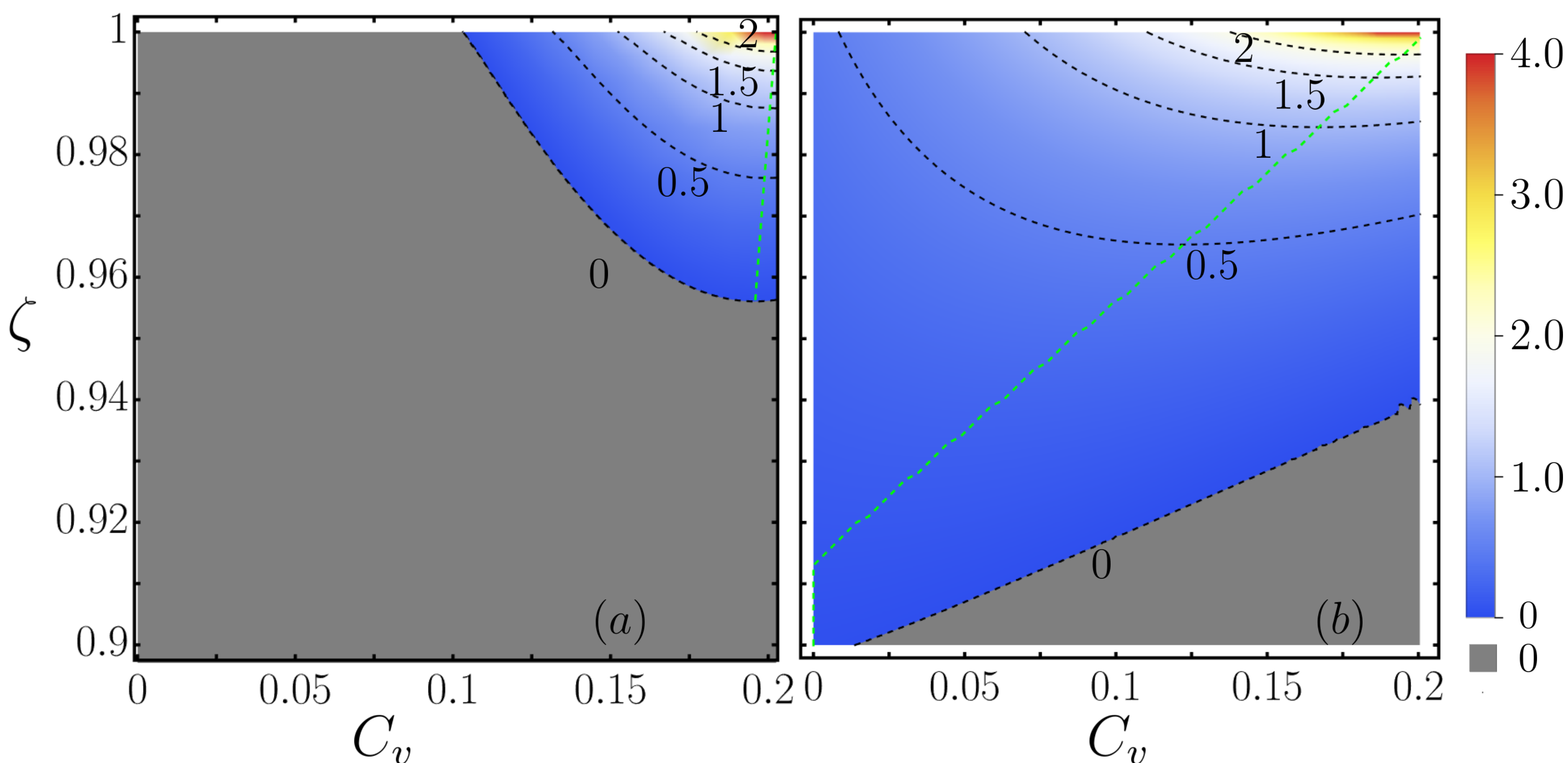}
    \caption{Contour of quantum information rate of microwave-to-optical transduction with $C_g=0.1$ and $n_{\rm{in}}=0.1$. (a) Quantum capacity lower bound of direct conversion. (b) Quantum capacity lower bound of teleportation. The green dashed line corresponds to optimum squeezing $C_v$. The boundary of zero is obtained by rate $\lesssim 0.01$ due to finite numerical precision. }
    \label{fig:contour2_Cg01}
\end{figure}
In the low cooperativity region of $C_g=0.1$,
Fig.~\ref{fig:line_combined} illustrates the quantum information rate and the effective noise $N$ versus the squeezing level $C_v$ for microwave-to-optical transduction and optical-to-microwave transduction. Both the lower and upper bounds of the direct conversion and teleportation channels are plotted. With chosen $\zeta$, microwave thermal noise $n_{\rm in}$ and cooperativity $C_g$,
for direct conversion, the capacity is directly determined by $C_v$; while for teleportation, we optimize the parameters $\kappa_q$ and $\kappa_p$ to maximize the lower bound and calculate the upper bound for each value of $C_v$. 
When the squeezing is small, we observe that the teleportation channel exhibits a non-zero capacity compared to direct conversion. The capacity lower bound of the teleportation channel surpasses the capacity upper bound of the direct conversion channel when $C_v<0.15$. For $C_v>0.15$, the capacity region (specified by lower and upper bounds) of the two approaches overlaps. To compare their performances, we manually set $\eta_{\rm{TP}}=\kappa_q\kappa_p=\eta_{\rm{DC}}$ so that teleportation approach and direct conversion scheme have identical transmissivity or gain, and compare the noise being mixed in. Remarkably, when $C_v>0.17$, we observe that $N_{\rm{TP}}<N_{\rm{DC}}$, indicating that the teleportation approach can achieve the same transmissivity of direct conversion with less noise. Thus, the teleportation approach outperforms direct conversion overall.

\begin{figure}[t]
    \centering
    \includegraphics[width=0.45\textwidth]{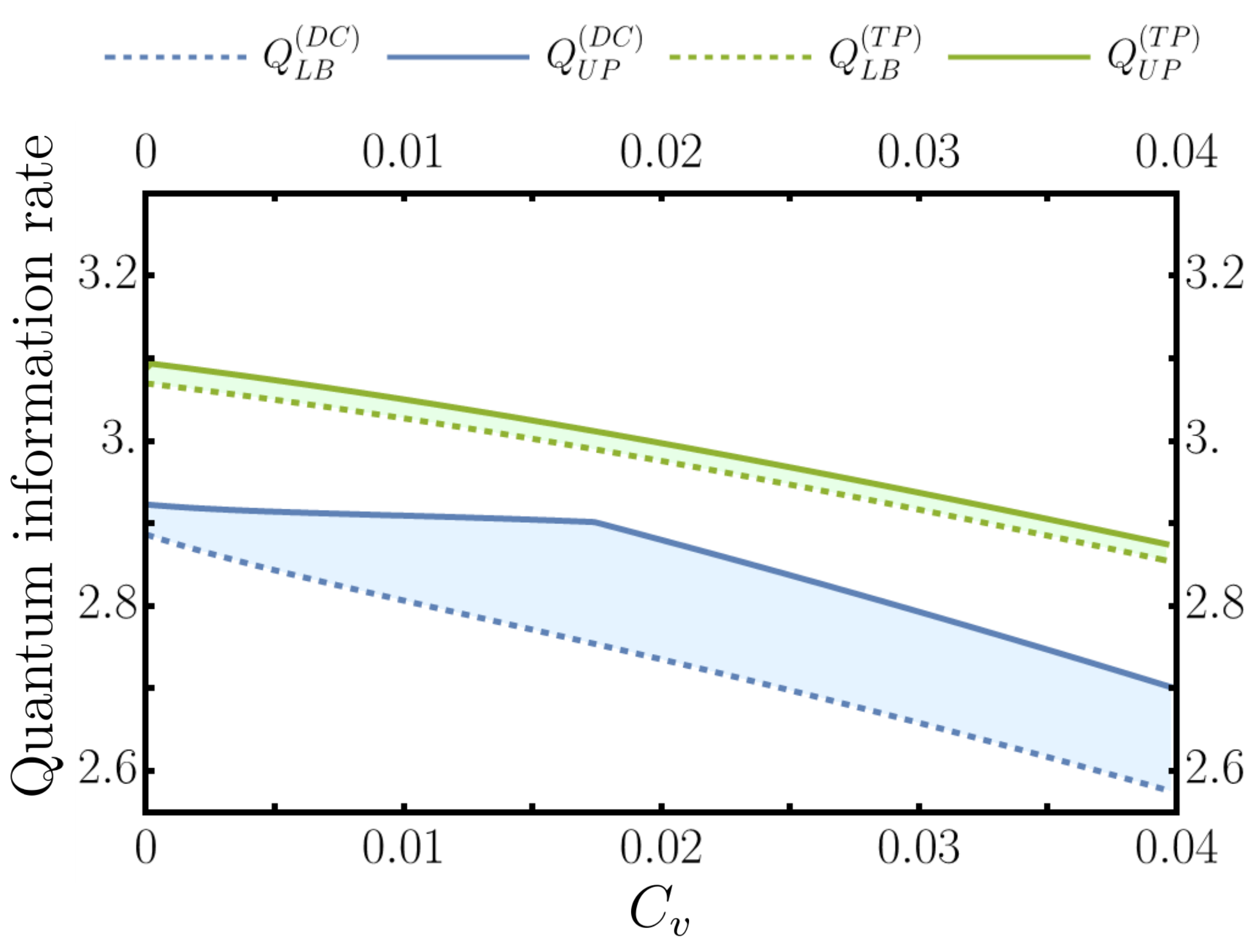}
    \caption{Capacity of microwave-to-optical transduction versus level of squeezing $C_v$ for extraction coefficiencies $\zeta_{\rm{m}}=\zeta_{\rm{o}}=0.975$, microwave thermal noise $n_{\rm{in}}=0.1$ and cooperativity $C_g=0.6$. Blue indicates direct conversion approach and green indicates teleportation approach. Solid and dashed curves indicate upper and lower bounds.
        \label{fig:capacityLine_Cg_06}
    }
\end{figure}

To provide a comprehensive comparison between the two approaches in the low cooperativity region, Fig.~\ref{fig:contour2_Cg01} shows the contour plots of the lower bound of the quantum information rate versus the extraction efficiency $\zeta$ and squeezing level $C_v$ at a fixed cooperativity of $C_g=0.1$. By comparing both approaches, we observe that the teleportation exhibits a larger region of non-zero rate and requires less optimum squeezing $C_v$ than the direct conversion approach for the same extraction coefficient $\zeta$, when $C_g=0.1$. 
Significantly, even without the help of any squeezing, the teleportation-based approach achieves a non-zero rate, as long as $\zeta\gtrsim 0.9$. In contrast, the direct conversion approach requires $\zeta\gtrsim 0.96$ to have anon-zero rate, with the help of optimal squeezing.

To maximize the rate for each extraction efficiency, we select the squeezing $C_v$ that maximizes the lower bound of the capacity while fixing $\zeta$, as indicated by the green dashed line. The advantage of the teleportation approach over the direct conversion approach decreases as $\zeta$ approaches unity. However, in the direct conversion approach, as shown in Fig.~\ref{fig:contour2_Cg01} (a), it requires large squeezing. While in the teleportation approach, as shown in Fig.~\ref{fig:contour2_Cg01} (b), the optimum squeezing increases linearly with the extraction coefficients $\zeta$. 

Before concluding this section, we point out one caveat of the rate comparison in Fig.~\ref{fig:contour2_Cg01}. Because we are comparing the lower bounds of both protocols, we cannot guarantee that the actual rate of the teleportation-based scheme is higher than the direct conversion scheme, similar to the situation in Ref.~\cite{zhong2022quantum}. However, in the presence of a noisy channel, achievable rates in the near-term are much closer to the lower bound when the channel is noisy~\cite{noh2018quantum}, despite the exact capacity is unknown. Moreover, in Fig.~\ref{fig:line_combined}(b) we have demonstrated the advantage from noise comparison.

\subsubsection{High cooperativity}
\label{subsection:practical high cooperativity}
When the cooperativity $C_g$ is large and extraction efficiencies are non-unity, increasing the squeezing $C_v$ does not enhance the communication rate. For instance, when $C_g=0.6$, both the upper and lower bounds decrease with $C_v$, as illustrated in Fig.~\ref{fig:capacityLine_Cg_06}. In this case, squeezing dose not help transduction. At the same time, the teleportation-based approach has strictly higher information rates, as the true rate lies within the green region for teleportation, strictly higher than the true rate within the blue region for direct conversion. 

\begin{figure}[t]
    \centering
    \includegraphics[width=0.48\textwidth]{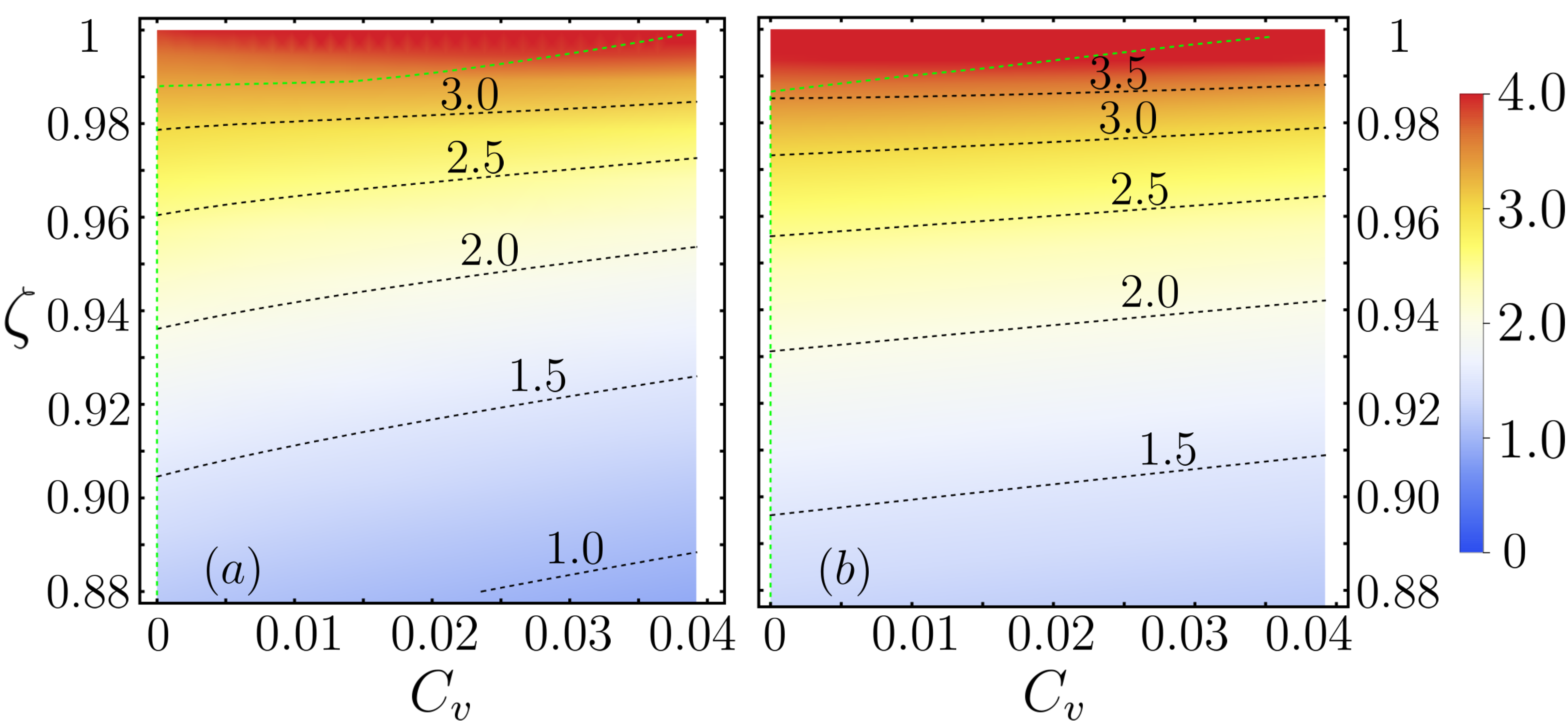}
    \caption{Contour of quantum information rate of microwave-to-optical transduction with $C_g=0.6$ and $n_{\rm{in}}=0.1$. (a) Lower bound of direct conversion. (b) Lower bound of teleportation. The green dashed line corresponds to optimum squeezing $C_v$. }
    \label{fig:contour2_Cg06}
\end{figure}

To gain further insights into the influence of extraction efficiency in the high cooperativity region,  
we present contour plots of the capacity versus $C_v$ and $\zeta$ contour when $C_g=0.6$ in Fig.~\ref{fig:contour2_Cg06}. For each fixed $\zeta$, we indicate the optimal squeezing level $C_v$ to maximize the rate in green dashed lines. In both approaches, we see that squeezing will not help unless the extraction efficiency is very close to unity. At the same time, the teleportation scheme still has advantage over the direct conversion in the maximum rates, despite the difference being smaller at large cooperativity.

\section{Fidelity comparison in state transduction}
\label{section:fidelity comparison}

\begin{figure}[t]
    \centering
    \includegraphics[width=0.48\textwidth]{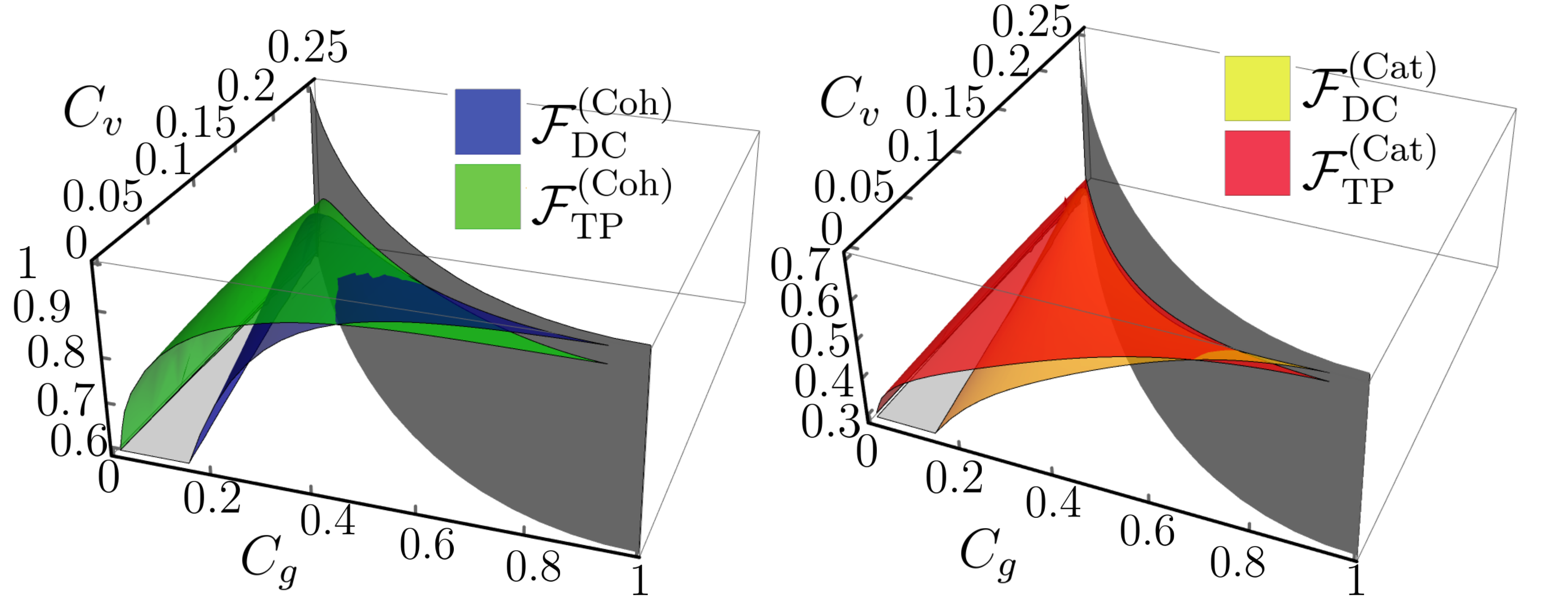}
    \caption{Fidelity of (a) coherent state (b) cat state after transduction when $\zeta_{\rm{o}}=\zeta_{\rm{m}}=0.95$ and $n_{\rm{in}}=0.1$. We have chosen the amplitude $\alpha=2$. Green: teleportation scheme. Blue: direct conversion scheme. Dark gray surface: boundary of the stable parameter region.}
    \label{fig:fidelity_coh}
\end{figure}

In previous sections, we compare the transduction performance between direct conversion and teleportation via quantum information rate analyses. Although being an accurate fundamental quantifier, quantum information rate in general requires specific encoding and decoding strategies to achieve, which can be challenging to realize in the near term. To compare the performance of the two approaches implemented in near-term systems, we consider the fidelity between the input and output states when transducing a practical quantum state. Here, we consider a coherent state $\ket{\alpha}$ and a cat state $\ket{cat_+}\propto(\ket{\alpha}+\ket{-\alpha})$ as examples, which are useful for quantum communication and computation. The Uhlmann fidelity $
\mathcal{F}\left(\hat{\rho},\hat{\sigma}\right)=\left({\rm tr}\sqrt{\sqrt{\hat{\rho}}\hat{\sigma}\sqrt{\hat{\rho}}}\right)^2
$~\cite{uhlmann1976transition,jozsa1994fidelity} measures the similarity between the input state $\hat{\rho}$ and output $\hat{\sigma}$ state, with unity value for identical input and output states in an ideal transduction system. When the input state is a pure state, which is our case for coherent states or cat states, the fidelity can be calculated explicitly by directly integrating Wigner functions as shown in Appendix~\ref{app:fidelity}. Again, in the teleportation scheme, we optimize $\kappa_q$ and $\kappa_p$ to maximize the fidelity of the output state.  

Fig.~\ref{fig:fidelity_coh} shows the fidelity comparison versus different values of cooperativity $C_g$ and squeezing levels $C_v$. The dark gray surface shows the boundary of the stable parameter region, below which the transduction system can operate. We see that in most of the visible parameter region, the fidelity of the teleportation-based approach (green) is higher than that of the direct conversion approach (blue). Only in the region with extremely high cooperativity, direct conversion demonstrates a slightly higher fidelity, when both fidelities are very close to unity.

\section{Conclusion and discussion}
To conclude, we analyze the role of single-mode squeezing in teleportation-based and direct conversion transduction via quantum information rates and fidelity of state conversion. Assuming ideal extraction efficiency, higher squeezing guarantees higher information rates for both protocols, while the teleportation-based approach still has advantage over the direct conversion approach. For non-ideal extraction efficiency, the teleportation-based approach provides an overall better performance. Even with the help of optimal squeezing, the direct conversion approach is much less tolerant against low extraction efficiencies. 

\begin{acknowledgements}
This work is supported by the National Science Foundation CAREER Award CCF-2142882, Office of Naval Research Grant No. N000142312296, Defense Advanced Research Projects Agency (DARPA) under Young Faculty Award (YFA) Grant No. N660012014029 and Cisco Systems, Inc.
\end{acknowledgements}

\appendix
\section{Single-mode Gaussian channel}
\label{app:gaussian_channel}
A succinct overview of Gaussian quantum information can be found in Appendix A of our prior work~\cite{wu2021deterministic}, while a full review of the topic can be found in Ref.~\cite{Weedbrook2012}. Here, we provide a brief summary. 

To describe an $M$-mode continuous-variable system, it is useful to define quadrature operators as  $\hat{q}_k \equiv (\hat{a}_k+\hat{a}^\dagger_k)/\sqrt{2}$, $\hat{p}_k \equiv i (\hat{a}^\dagger_k-\hat{a}_k)/\sqrt{2}$, where $\hat{a}_k$'s are the annihilation operators of the $k$th mode and we have chosen the unit of $\hbar=1$.
A quantum state $\hat{\rho}$ is Gaussian if its Wigner function $W(\bx)$ has a Gaussian form, which can be fully characterized by its $2M$-dimensional mean vector $\Bar{\bx}$ and $2M$ by $2M$ covariance matrix $\mathbf{V}$:
\begin{align}
    & \Bar{\bx} \equiv \text{Tr}[\hat{\rho} \hat{\bx}],\\
    & {\bm V}_{ij} \equiv
    \frac{1}{2}\text{Tr}[\hat{\rho} \{\hat{x}_i-\bar{x}_i,\hat{x}_j-\bar{x}_j\}].
\end{align}

A single-mode ($M=1$) Gaussian channel transforms an arbitrary single-mode Gaussian state $\hat{\rho}(\Bar{\bx},\mathbf{V})$ by the linear and bilinear transform of the mean and covariance matrix:
\begin{equation}
    \Bar{\bx}\rightarrow \mathbf{T} \Bar{\bx}+\mathbf{d}, \qquad \mathbf{V} \rightarrow \mathbf{T}\mathbf{V}\mathbf{T}^{\top}+\mathbf{N},
\end{equation}
where $\mathbf{d}$ is a $2$-dimensional real displacement vector,  $\mathbf{T}$ and $\mathbf{N}$ are $2$-by-$2$ real matrices that satisfy the condition:
\begin{equation}
    \mathbf{N}+i \mathbf{\Omega}-i \mathbf{T}\mathbf{\Omega}\mathbf{T}^\top \ge 0,
    \label{app:single_mode_Gaussian_condition}
\end{equation}
with $\mathbf{\Omega}=\begin{pmatrix}
    0 & 1\\
    -1 & 0
\end{pmatrix}$.
We are interested in Gaussian channels that can be reduced to an additive Gaussian noise channel, a thermal-loss channel, or a thermal-amplification channel via Gaussian unitaries, where
the transmissivity $\eta$ and the noise being mixed in $N$ are given by
\begin{align}
    & \eta = \det\mathbf{T},\\
    & N=\left(\Bar{n}+\frac{1}{2}\right)|1-\eta |=\sqrt{\det\mathbf{N}}.
\end{align}
Here $\det\mathbf{A}$ denotes the determinant of a matrix $\mathbf{A}$.
The quantum capacity lower bound of a single-mode Gaussian channel with transmissivity $\eta$ and thermal noise $\Bar{n}$ is given by~\cite{holevo2001evaluating}:
\be 
Q_{\rm {LB}}(\eta,\Bar{n}) \equiv  \max\left[\log_2\left(\frac{\eta}{|1-\eta|}\right)-g\left(\Bar{n}\right),0\right],
\ee
where the function
$ 
g(x)=(x+1)\log_2(x+1)-x\log_2 x
$
is the von Neumann entropy of a thermal state with mean occupation number $x$. When $\eta\to 1$, the channel becomes an additive noise channel with additive noise $N_{\rm {add}}=(1-\eta)\bar n$. The capacity lower bound is then given by
\be 
Q_{\rm LB}(\eta,\Bar{n})=-\log_2(N_{\rm{add}})-1/\ln(2).
\ee 
We will use upper bounds that have been derived from two-way assisted quantum capacity~\cite{pirandola2017fundamental} and the degradable extensions~\cite{fanizza2021estimating}. Combining these results, we have
\begin{equation}
Q_{\rm UP}=\min[Q_{\rm PLOB}(\eta,\Bar{n}), Q_{\rm DE}(\eta,\Bar{n}) ].
\end{equation}
The functions are
\begin{widetext}
\begin{align}
 Q_{\rm PLOB}(\eta, \Bar{n}) &\equiv 
    \begin{cases}
         \max\left[-\log_2\left[\left(1-\eta\right)\eta^{\Bar{n}}\right]-g\left(\Bar{n}\right),0\right] 
         & \mbox{if $\eta< 1$};\\
        \max\left[\log_2\left(\frac{\eta^{\Bar{n}+1}}{\eta-1}\right)-g(\Bar{n}),0\right] 
        & \mbox{if $\eta>1$},\\ 
        \log_2\left(1/N_{\rm add}\right)-1/\ln(2)+N_{\rm add}/\ln(2).
        & \mbox{if $\eta=1, (1-\eta)\Bar{n}\to N_{\rm add}$},\\
    \end{cases}\\
 Q_{\rm DE}(\eta, \Bar{n}) & \equiv 
    \begin{cases}
         \max\left[\log_2\left(\frac{\eta}{1-\eta}\right)+h\left[\left(1-\eta\right)\left(2\Bar{n}+1\right)+\eta\right]- h\left[\eta\left(2 \Bar{n} +1\right)+1-\eta\right],0\right] 
         & \mbox{if $\eta< 1$};\\
        \max\left[
        \log_2\left(\frac{1}{\left(\eta-1\right)\Bar{n}}\right)-1/\ln{2}+2h\left(\sqrt{1+\left(\eta-1\right)^2 \Bar{n}^2}\right),0
        \right]
        & \mbox{if $\eta>1$};\\
         \max\left[
        \log_2\left(\frac{1}{N_{\rm add}}\right)-1/\ln{2}+2h\left(\sqrt{1+N_{\rm add}^2}\right),0
        \right]
        & \mbox{if $\eta=1, (1-\eta)\Bar{n}\to N_{\rm add}$}.
    \end{cases}
\end{align}
\end{widetext}
The function $h(x)=g(\frac{x-1}{2})$.

\section{Quantum channel model of direct conversion}
\label{app:DC_channel}
In the direct conversion scheme, the interaction between optical mode $\hat{b}$ and microwave mode $\hat{m}$ can be modeled as a beam-splitter-like Hamiltonian given by
\begin{align}
    \hat{H}_{\rm{BS}}&=g (\hat{b}^{\dagger}\hat{m}+\hat{b}\hat{m}^{\dagger}),
\end{align}
where the pumping photon number $N_p$ of optical mode $\hat{a}$ has been absorbed into $g$. When squeezing is involved, the total Hamiltonian can be written as 
\begin{align}
    \hat{H}_{\rm{DC}}/\hbar=& \omega_{\rm{o}} \hat{b}^{\dagger}\hat{b} + \omega_{\rm{m}} \hat{m}^{\dagger}\hat{m} +  \hat{H}_{\rm{BS}} + \hat{H}_{\rm{MS/OS}},
    \label{eq:app_Hamiltonian_DC}
\end{align}
where $\hat{H}_{\rm{MS}}$ and $\hat{H}_{\rm{OS}}$ model the microwave squeezing and optical squeezing, respectively. They can be expressed as
\begin{align}
    \hat{H}_{\rm{MS}}&= v(e^{i \theta }\hat{m}^2+e^{-i \theta } \hat{m}^{\dagger 2}),
    \label{eq:app_squeezingHamiltonian_MS}\\
    \hat{H}_{\rm{OS}}&= v(e^{i \theta }\hat{b}^2+e^{-i \theta } \hat{b}^{\dagger 2}),
    \label{eq:app_squeezingHamiltonian_OS}
\end{align}
where $v$ and $\theta$ are real numbers describing the squeezing level and rotation, respectively.
\subsection{Stable condition}
\label{app:DC_stable_condition}
Let's consider the stability condition of the transduction system, which can be derived via the Laplace analysis as in Ref.~\cite{PhysRevA.84.043845}. The quantum Langevin equations can be obtained from the Hamiltonian in Eq.~\eqref{eq:app_Hamiltonian_DC}. For example, for single-mode squeezing in microwave, they are
\begin{subequations}
\begin{align}
    \dv{\hat{b}(t)}{t} &= i\Delta_{\rm{o}} \hat{b} - i g \hat{m}-\frac{\gamma_{\rm{o}}}{2} \hat{b} +\sqrt{\gamma_{pc}} \hat{b}_{\rm{in}} + \sqrt{\gamma_{pi}} \hat{b}^{(i)},  \\
    \dv{\hat{m}(t)}{t} &= i \Delta_{\rm{m}} \hat{m} -ig \hat{b} - 2ve^{-i \theta} \hat{m}^\dagger - \nonumber \\ &\qquad \frac{\gamma_{\rm{m}}}{2} \hat{m} +\sqrt{\gamma_{ec}} \hat{m}_{\rm{in}} + \sqrt{\gamma_{ei}} \hat{m}^{(i)},\\
   \hat{b}_{\rm{out}}&=-\sqrt{\gamma_{pc}} \hat{b} + \hat{b}_{\rm{in}},\\
   \hat{m}_{\rm{out}}&=-\sqrt{\gamma_{ec}} \hat{m} + \hat{m}_{\rm{in}},
\end{align}
\end{subequations}
where $\hat{b}$, $\hat{m}$ are time-dependent operators. The terms $\hat{b}_{\rm{in}}$ and $\hat{b}^{(i)}$ correspond to input and intrinsic loss of optical mode with $\gamma_{\rm{o}}=\gamma_{pc}+\gamma_{pi}$. $\hat{m}_{\rm{in}}$ and $\hat{m}^{(i)}$ are input and intrinsic loss of microwave mode with $\gamma_{\rm{m}}=\gamma_{ec}+\gamma_{ei}$. The detunings of the optical and microwave modes are denoted by $\Delta_o$ and $\Delta_m$, respectively. The Laplace transform of function $f(t)$ is defined as
\begin{align}
    & \tilde{f}(s)= \int_0^\infty \diff{t} f(t) \exp(-st),
\end{align}
which allows the replacement rule $\dv{f(t)}{t} \rightarrow s \tilde{f}(s) - f(0)$.
By taking the Laplace transform and assuming zero detuning, we obtain the following quantum Langevin equations in the complex domain,
\begin{align}
    &0=\mathbf{G}_{\rm{DC},MS/OS}(s)\tilde{\mathbf{b}} + \tilde{\mathbf{b}}(0) + \mathbf{K} \tilde{\mathbf{b}}_{\rm{in}}, \\
    &  \mathbf{\tilde{b}}_{\rm{out}}(s) = -\mathbf{K}^{\rm{T}}\tilde{\mathbf{b}}+\tilde{\mathbf{b}}_{\rm{in}},\\
    &
    \mathbf{\tilde{b}}_{\rm{out}}(s)=\mathbf{F}(s) \mathbf{\tilde{b}}(0) + \mathbf{S}(s) \mathbf{\tilde{b}}_{\rm{in}}(s).
\end{align}
The vectors of operators are defined by following
\begin{align}
    &\mathbf{\tilde{b}}=\left(\tilde{b}, \tilde{b}^\dagger, \tilde{m}, \tilde{m}^\dagger \right)^{\rm T},\\
    &\tilde{\mathbf{b}}_{\rm{in}}=(\tilde{b}_{\rm {in}}, \tilde{b}^\dagger_{\rm {in}},\tilde{b}^{(i)}, \tilde{b}^{\dagger(i)}, \tilde{m}_{\rm {in}}, \tilde{m}^\dagger_{\rm {in}},\tilde{m}^{(i)}, \tilde{m}^{(i)\dagger}  )^{\rm T},
\end{align}
where the tilde version of operators denote the operators' Laplace transforms.
And 
\begin{align}
    &\mathbf{G}_{\rm{DC,MS}}(s)=\begin{pmatrix}
    -\frac{\gamma_{\rm{o}}}{2}-s & 0 & -i g & 0\\
    0 & -\frac{\gamma_{\rm{o}}}{2}-s^* & 0 & i g\\
     -i g & 0 & -\frac{\gamma_{\rm{m}}}{2}-s & -2 i v e^{-i \theta}\\
    0 & i g & 2 i v e^{i \theta} & -\frac{\gamma_{\rm{m}}}{2}-s^*
    \end{pmatrix},\\
    &\mathbf{G}_{\rm{DC,OS}}(s)=\begin{pmatrix}
    -\frac{\gamma_{\rm{o}}}{2}-s & -2 i v e^{-i \theta} & -i g & 0\\
    2 i v e^{i \theta} & -\frac{\gamma_{\rm{o}}}{2}-s^* & 0 & i g\\
     -i g & 0 & -\frac{\gamma_{\rm{m}}}{2}-s & 0 \\
    0 & i g & 0 & -\frac{\gamma_{\rm{m}}}{2}-s^*
    \end{pmatrix},\\
    &\mathbf{K}=\begin{pmatrix}
    \sqrt{\gamma_{pc}}& 0 &\sqrt{\gamma_{pi}} & 0 & 0 & 0 & 0 & 0 \\
    0 & \sqrt{\gamma_{pc}}& 0 &\sqrt{\gamma_{pi}} & 0 & 0 & 0 & 0\\
    0 & 0 & 0 & 0 & \sqrt{\gamma_{ec}}& 0 &\sqrt{\gamma_{ei}} & 0\\
    0 & 0 & 0 & 0 & 0 & \sqrt{\gamma_{ec}}& 0 &\sqrt{\gamma_{ei}}
    \end{pmatrix},\\
&  \mathbf{F}(s) = \mathbf{K}^{\rm{T}}\mathbf{G}_{\rm{DC,MS/OS}}(s)^{-1}.
\end{align}    

To analyze the stability of the system, we need to ensure that the poles of the transfer function $\mathbf{F}(s)$ are located in the left half-plane of the complex plane. This can be achieved by examining the roots of the equation $\det \mathbf{G}_{\text{DC}}(\alpha) = 0$, where $\alpha$ is a real variable. The condition for stability is that the largest root of this equation must be negative. After solving the equation, we obtain the stability condition:
\begin{equation}
\label{eq:app_DC_stable}
    1+C_g-2\sqrt{C_v}>0,
\end{equation}
where $C_v={4v^2}/{\gamma_{\rm{m/o}}^2}$ for microwave or optical squeezing respectively and $C_g=4g^2/(\gamma_{\rm{m}}\gamma_{\rm{o}})$ characterizes the strength of the beam-splitter interaction.

\subsection{Output field}
\label{app:DC_output_field}
The input-output relations between the microwave/optical input and the optical/microwave output can be derived using the quantum Langevin equations in the frequency domain.  We begin by considering the general relation for the cavity modes, which includes the input modes $\bbold_{\rm {in}}$, output modes $\bbold_{\rm {out}}$ and cavity modes $\bbold$ in frequency domain. They are defined as follows
\begin{align}
    &\bbold=(\hat{b}, \hat{b}^\dagger, \hat{m}, \hat{m}^\dagger  )^{\rm T},\\
    \label{eq:App-field-notation1}
    &\bbold_{\rm {in}}=(\hat{b}_{\rm {in}}, \hat{b}^\dagger_{\rm {in}},\hat{b}^{(i)}, \hat{b}^{\dagger(i)}, \hat{m}_{\rm {in}}, \hat{m}^\dagger_{\rm {in}},\hat{m}^{(i)}, \hat{m}^{\dagger(i)}  )^{\rm T},\\
    \label{eq:App-field-notation2}
    &\bbold_{\rm {out}}=(\hat{b}_{\rm {out}}, \hat{b}^\dagger_{\rm {out}},\hat{b}^{(i)}, \hat{b}^{\dagger(i)}, \hat{m}_{\rm {out}}, \hat{m}^\dagger_{\rm {out}},\hat{m}^{(i)}, \hat{m}^{\dagger(i)}  )^{\rm T}.
\end{align}
Replacing $s$ by $i \omega$ in the quantum Langevin equations in complex domain, the output fields can be derived by solving a set of Heisenberg-Langevin equations in the frequency domain with using the input-output relations\cite{zhong2020proposal,cui2021high}
\begin{align}
    &0=\mathbf{G}(i\omega)\bbold + \mathbf{K}\bbold_{\rm {in}},\\
    &\bbold_{\rm {out}}=-\mathbf{K}^{\rm T}\bbold+\bbold_{\rm {in}}.
    \label{eq:input-output-Langevin}
\end{align}
Here we utilize the matrix notation to represent the dynamics in the Fourier domain. The output field is then given by
\begin{align}
     \bbold_{\rm {out}}& = (\mathbf{K}^{\rm T}\mathbf{G}^{-1}(i\omega)\mathbf{K}+\mathbf{I}_{8}) \bbold_{\rm {in}}  \equiv \mathbf{S}(i\omega)\bbold_{\rm {in}}.
     \label{eq:input-output-Fourier}
\end{align}
For the case of zero detuning when $\omega=0$, we use the shorter notations $\mathbf{G}$ and $\mathbf{S}$. 
To transform the input-output relatoin from annihilation and creation operators $(\hat{a}, \hat{a}^{\dagger})$ to quadrature operators $\hat{x}=(\hat{q}_a,\hat{p}_a)$, we introduce the transformation matrix: 
\begin{align}
    & \mathbf{Q}=\mathbf{I}_4 \otimes \frac{1}{\sqrt{2}}\begin{pmatrix}
    1 & 1\\
    -i & i
    \end{pmatrix}.
\end{align}
We obtain the relation between the input and output fields as
\begin{align}
    &  \hat{\mathbf{x}}_{\rm{out}} =  \mathbf{S}_{\mathbf{x}}\hat{\mathbf{x}}_{\rm{in}},\\
    &   
    \mathbf{S}_{\mathbf{x}} = \mathbf{Q} \mathbf{S} \mathbf{Q}^{-1}, \label{eq:input-output-x}\\
    &\mathbf{V}_{\rm{out}} = \mathbf{S}_{\mathbf{x}} \mathbf{V}_{\rm{in}}\mathbf{S}_{\mathbf{x}}^\top
    \label{eq:app_Vin_Vout},
\end{align}
where $\hat{\mathbf{x}}_{\rm{in}}$ and $ \hat{\mathbf{x}}_{\rm{out}}$ represent the quadrature operators of the input and output fields, and $\mathbf{V}_{\rm{out}}$ and $\mathbf{V}_{\rm{in}}$ are the covariance matrices of the output and input fields, respectively.

\subsection{Transmissivity and noise}
\label{app:DC_channel_eta_n_}
In the previous section, we introduced the notations for input and output fields and derived the general relation for a multi-mode Gaussian channel as Eq.~\eqref{eq:input-output-x} for zero detunings. The single-mode transduction channel can be obtained from the reduced channel of this general Gaussian channel. For example, by tracing out the other modes at the output, we obtain the reduced channel that describes the microwave-to-optical transduction:
\begin{align}
    \begin{pmatrix}
    \hat{q}_{\rm{o}} \\ 
    \hat{p}_{\rm{o}}
    \end{pmatrix}&=\mathbf{S}^{<\{1,2\},\{5,6\}>}_{\mathbf{x}} \begin{pmatrix}
    \hat{q}_{\rm{m}} \\ 
    \hat{p}_{\rm{m}}
    \end{pmatrix}  \equiv \mathbf{T} \begin{pmatrix}
    \hat{q}_{\rm{m}} \\ 
    \hat{p}_{\rm{m}}
    \end{pmatrix},
\end{align}
where $<\{1,2\},\{5,6\}>$ presents the row $1,2$ and column $5,6$ block of the matrix $\mathbf{S}_{\mathbf{x}}$. The transmissivity of the microwave-to-optical transduction can be calculated as:
\begin{equation}
\eta_{\rm{DC,MS}}^{\rm m\veryshortrightarrow o}=\det\mathbf{T}=\frac{4C_g \zeta_{\rm{o}}  \zeta_{\rm{m}}}{(1+C_g)^2-4C_v}.
\end{equation}
The noise matrix for the microwave-to-optical transduction can be calculated as:
\begin{align}
    \mathbf{N}_{\rm{DC,MS}}^{\rm m\veryshortrightarrow o}=\mathbf{S}^{<\{1,2\},\{1,2,3,4,7,8\}>}_{\mathbf{x}}\mathbf{V}_{\rm{in}}\mathbf{S}^{<\{1,2\},\{1,2,3,4,7,8\}>\rm{T}}_{\mathbf{x}},
    \label{eq:Noise_mixed_in_appendix}
\end{align}
where we assume the microwave thermal noise $n_{\rm{in}}$ and zero optical thermal loss resulting in the input covariance matrix:
\begin{equation}
\mathbf{V}_{\rm{in}}=\text{Diag}\left(\frac{1}{2},\frac{1}{2},\frac{1}{2},\frac{1}{2},\frac{1}{2}+n_{\rm{in}},\frac{1}{2}+n_{\rm{in}}\right).
\end{equation}
We can then calculate the noise $\Bar{n}_{\rm{DC,MS}}^{\rm m\rightarrow o}$ and the noise mixed in $N_{\rm{DC,MS}}^{\rm m\veryshortrightarrow o}$ through the following expression,
\begin{align}
    N_{\rm{DC,MS}}^{\rm m\veryshortrightarrow o}&=\left(\frac{1}{2}+ \Bar{n}_{\rm{DC,MS}}^{\rm m\veryshortrightarrow o}\right)|1-\eta_{\rm{DC,MS}}^{\rm m\veryshortrightarrow o}| \nonumber\\
    &=\sqrt{\det\mathbf{N}_{\rm{DC,MS}}^{\rm m\veryshortrightarrow o}}.
\end{align}
This leads to the result:
\begin{equation}
    N_{\rm{DC,MS}}^{\rm m\veryshortrightarrow o}=\frac{\sqrt{\left[A_-^2+4C_g\zeta_{\rm{o}}B_+\right]\left[A_+^2+4C_g\zeta_{\rm{o}}B_-\right]}}{2\left[(1+C_g)^2-4C_v\right]},
\end{equation}
where
\begin{align}
    &A_-=1+C_g-2\sqrt{C_v},\\
    &A_+=1+C_g+2\sqrt{C_v},\\
    &B_+=(1+2n_{\rm{in}})(1-\zeta_{\rm{m}})-1+2\sqrt{C_v},\\
    &B_-=(1+2n_{\rm{in}})(1-\zeta_{\rm{m}})-1-2\sqrt{C_v}.
\end{align}
The noise is independent of the phase rotation angle $\theta$ of the squeezing. From the above calculations we find that if $\zeta_{\rm{o}}=\zeta_{\rm{m}}=1$, we have
\begin{align}
    \Bar{n}_{\rm{DC,MS}}^{\rm m\veryshortrightarrow o}=0.
    \label{eq:ideal_nDC}
\end{align}
In this case, the capacity lower and upper bounds coincide since the thermal loss $\Bar{n}=0$. 
Via the theory framework above, we can analyze the optical to microwave direction with microwave or optical squeezing, and we find that the transmissivities $\eta_{\rm{DC,MS/OS}}^{\rm m\veryshortrightarrow o}$ and $\eta_{\rm{DC,MS/OS}}^{\rm o\veryshortrightarrow m}$ are the same for microwave-to-optical transduction and optical-to-microwave transduction with the microwave and optical squeezing coefficient $C_v$ defined as $4v^2/\gamma_{\rm{m}}^2$ and $4v^2/\gamma_{\rm{o}}^2$, respectively. However, the noise being mixed in, $N_{\rm{DC,MS/OS}}^{\rm m\veryshortrightarrow o}$ and $N_{\rm{DC,MS/OS}}^{\rm o\veryshortrightarrow m}$, are generally different. The expressions of $N_{\rm{DC}}$ are lengthy and will not be presented here. But when $\zeta_{\rm{o}}=\zeta_{\rm{m}}=1$, for all direct conversion channels we have
\begin{align}
    \Bar{n}_{\rm{DC}}=0.
\end{align}

\section{Quantum channel model for teleportation-based transduction}
\label{app:TP_channel}
In the teleportation approach, the optical mode $\hat{b}$ in Eq.~\eqref{eq:Interaction_hamiltonian_MS} is pumped to generate a two-mode squeezing interaction between optical mode $\hat{a}$ and microwave mode $\hat{m}$. Thus, the total Hamiltonian $\hat{H}_{\rm TP}$ is
\begin{align}
    \hat{H}_{\rm{TP}}/\hbar&= \omega_{\rm{o}} \hat{a}^{\dagger}\hat{a} + \omega_{\rm{m}} \hat{m}^{\dagger}\hat{m} + \hat{H}_{\rm{TMS}}+\hat{H}_{\rm{MS/OS}},\\
    \hat{H}_{\rm{TMS}}&=g (\hat{a}^{\dagger}\hat{m}^{\dagger}+\hat{a}\hat{m}),
    \label{eq:app_Hamiltonian_TP}
\end{align}
where $\hat{H}_{\rm{MS}}$ and $\hat{H}_{\rm{OS}}$ are single-mode squeezing in microwave and optical domain. In the following, we will provide the stable condition in the teleportation scheme, derive the covariance matrix of the entangled state and finally show that the input-output relation can be represented by a single-mode Gaussian channel.
\subsection{Stable condition}
\label{app:TP_stable_condition}
The stable condition can be derived in the same fashion as Section~\ref{app:DC_stable_condition}. With Hamiltonian given by Eq.~\eqref{eq:app_Hamiltonian_TP}, the quantum Langevin
equations are obtained in the teleportation scheme. Instead of the full equations, we directly provide the transition matrices as
\begin{align}
    \mathbf{G}_{\rm{TP,MS}}(s)&=
    \begin{pmatrix}
    -\frac{\gamma_{\rm{o}}}{2}-s & 0 & 0 & -i g\\
    0 & -\frac{\gamma_{\rm{o}}}{2}-s^* & i g & 0\\
    0 & - i g & -\frac{\gamma_{\rm{m}}}{2}-s & -2 i v e^{-i \theta}\\
    i g & 0 & 2 i v e^{i \theta} & -\frac{\gamma_{\rm{m}}}{2}-s^*
    \end{pmatrix},\\
    \mathbf{G}_{\rm{TP,OS}}(s)&=\begin{pmatrix}
    -\frac{\gamma_{\rm{o}}}{2}-s &  -2 i v e^{-i \theta} & 0 & -i g\\
    2 i v e^{i \theta} & -\frac{\gamma_{\rm{o}}}{2}-s^* & i g & 0\\
     0 & -ig & -\frac{\gamma_{\rm{m}}}{2}-s & 0\\
    ig & 0 & 0 & -\frac{\gamma_{\rm{m}}}{2}-s^*
    \end{pmatrix}.
\end{align}
The maximum root of the equation $\det \mathbf{G}_{\rm{TP}}(\alpha)=0$ has to be smaller then $0$ to enable a stable system. Thus, we have the stable condition for entanglement generation
\begin{equation}
    1-C_g-2\sqrt{C_v}>0,
\end{equation}
where $C_v=\frac{4v^2}{\gamma_{\rm{m/o}}^2}$ for microwave or optical squeezing respectively.

\subsection{Covariance matrix of the entangled state}
\label{app:TP_covariance_matrix}
The state generated from the cavity is a multi-mode Gaussian state as the Hamiltonian $\hat{H}_{\rm TP}$ is of second order. The relation between the input and output covariance matrix is the same as Eq.~\eqref{eq:app_Vin_Vout}, while we get the symplectic transform $\mathbf{S}_{\bx}$ from $\mathbf{G}_{\rm{TP}}$. There are two cases, single-mode squeezing in either microwave or optical mode. We assume vacuum for optical intrinsic mode and a thermal state with mean photon number $n_{\rm in}$ for microwave intrinsic mode. In both cases, the covariance matrix of the generated Gaussian state can be written in the following form
\begin{align}
    V_{\rm{om}}  \equiv \begin{pmatrix}
    \sigma_{\rm{o}} & \sigma_{\rm{om}}\\
     \sigma_{\rm{om}}^T & \sigma_{\rm{m}}
    \end{pmatrix},
    \label{eq:CM_TP}
\end{align}
where 
\begin{align}
\sigma_{\rm{o}}=\frac{1}{2}\begin{pmatrix}w_1-w_2 \sin{\theta}& w_2 \cos{\theta} \\
    w_2 \cos{\theta} & w_1+w_2 \sin{\theta}
    \end{pmatrix},
\\
\sigma_{\rm{m}}=\frac{1}{2}\begin{pmatrix}
u_1-u_2\sin{\theta} & -u_2\cos{\theta}\\
u_2\cos{\theta} & u_1+u_2\sin{\theta}
\end{pmatrix},
\\
\sigma_{\rm{om}}=\frac{1}{2}\begin{pmatrix}
-v_2\cos{\theta} & v_1+v_2\sin{\theta}
\\
v_1-v_2\sin{\theta} & -v_2\cos{\theta}
\end{pmatrix},
\end{align}
and the parameters
$w_1$, $w_2$, $u_1$, $u_2$, $v_1$ and $v_2$ are independent of $\theta$. 

We quantify the amount of entanglement with the reverse coherent information, which characterizes an achievable rate in the quantum communication. 
\begin{align}
    \mbox{RCI}^{\rm m\veryshortrightarrow o} &\equiv H(\hat{\rho}_{\rm{o}})-H(\hat{\rho}_{\rm{o,m}}),\\
    \mbox{RCI}^{\rm o\veryshortrightarrow m} &\equiv H(\hat{\rho}_{\rm{m}})-H(\hat{\rho}_{\rm{o,m}}),
\end{align}
where $H(\hat{\rho}_{\rm{m}})$, $H(\hat{\rho}_{\rm{o}})$ and $H(\hat{\rho}_{\rm{o,m}})$ are the von Neumann entropy of microwave mode, optical mode and total system, respectively, with $H(\hat{\rho})=-\tr \left(\hat{\rho}\log_2 \hat{\rho}\right)$. For Gaussian states, the entropy is a function of the symplectic eigenvalues of $\sigma_{\rm{m}}$, $\sigma_{\rm{o}}$ and $V_{\rm{om}}$. As they are independent of $\theta$, the entanglement of the state is independent of $\theta$ as well. So we take $\theta=\pi/2$, and apply $\pi/2$ rotation on microwave mode for further simplification. The covariance matrix can be simplified to a standard form
\begin{align}
V_{\rm{mo}} & = \frac{1}{2}
\begin{pmatrix}
u_1-u_2 & 0 & v_1+v_2 & 0   \\
0 & u_1+u_2 & 0 & v_1-v_2 \\
v_1+v_2 & 0 & w_1-w_2& 0\\
0 & v_1-v_2 & 0 & w_1+w_2
\end{pmatrix}\nonumber\\
& =  \frac{1}{2}
\begin{pmatrix}
u_q & 0 &v_q & 0   \\
0 & u_p & 0 & -v_p \\
v_q & 0 & w_q & 0\\
0 & -v_p & 0 & w_p
\end{pmatrix},
\label{eq:app_TP_CM_standard_form}
\end{align}
where the parameters are given in Eq.~\eqref{eq:TP_parameters_MS} for microwave squeezing. As for optical squeezing, they are
\begin{align}
    & u_{q/p}^\prime= 1+\frac{8\zeta_{\rm{m}}[C_g(1\pm\sqrt{C_v})+(1\pm2\sqrt{C_v})^2 n_{\rm{in}}(1-\zeta_{\rm{m}})]}{(1-C_g\pm2\sqrt{C_v})^2},\\
    & w_{q/p}^\prime=1+\frac{8\zeta_{\rm o}[C_g(1+n_{\rm{in}}(1-\zeta_{\rm{m}}))\mp \sqrt{C_v}]}{(1-C_g \pm 2\sqrt{C_v})^2},\\
    & v_{q/p}^\prime=\frac{4\sqrt{C_g \zeta_{\rm{m}} \zeta_{\rm{o}}}(1+C_g+2(1 \pm 2\sqrt{C_v})n_{\rm{in}}\left(1-\zeta_{\rm{m}} )\right)}{(1-C_g\pm 2\sqrt{C_v})^2},
    \label{eq:TP_parameters_OS}
\end{align}
where $C_v=4v^2/\gamma_{\rm{o/m}}^2$. When $\zeta_{\rm{m}}=\zeta_{\rm{o}}=1$, the symplectic eigenvalues of $V_{\rm{mo}}$ are $(1/2,1/2)$, regardless of microwave or optical squeezing. The RCI is determined by symplectic eigenvalues of $\sigma_{\rm{o}}$ and $\sigma_{\rm{m}}$,
\begin{equation}
    \mbox{RCI} =\frac{2s+1}{2}\log_2(\frac{2s+1}{2})-\frac{2s-1}{2}\log_2(\frac{2s-1}{2}),
\end{equation}
where 
\begin{align}
s=s_{\rm{o}}=s_{\rm{m}}=\sqrt{\frac{1}{4}+ \frac{4C_g (1+C_g)^2}{[(1-C_g)^2-4C_v]^2}}.
    \label{eq:Symplectic-eigenvalues}
\end{align}

\subsection{Single-mode Gaussian channel modeling}
\label{app:TP_Gaussian channel}
\begin{figure}[t]
    \centering
    \includegraphics[width=0.45\textwidth]{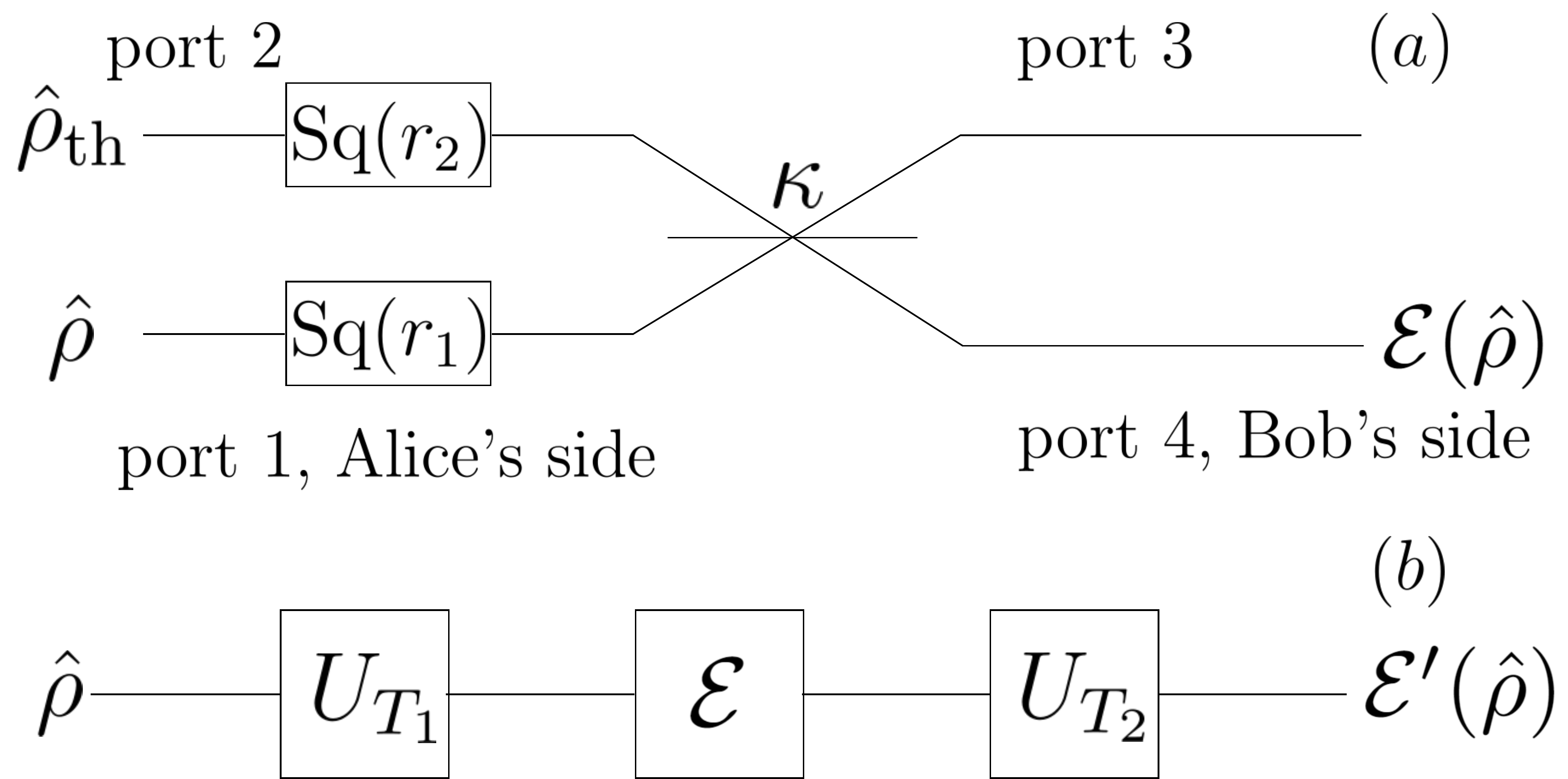}
    \caption{(a) The equivalent single-mode channel of quantum teleportation utilizing the Gaussian state with standard form in~\eqref{eq:app_TP_CM_standard_form}. (b) The single-mode Gaussian channel can be reduced to thermal-loss or thermal-amplification channel with corresponding pre-processing and post-processing.}
    \label{fig:channel1}
\end{figure}

Quantum teleportation based transduction scheme utilizes the optical-microwave entangled state generated from the electro-optical cavity to teleport the input state to the output state. Here, we show that the quantum teleportation utilizing the Gaussian state with covariance matrix in Eq.~\eqref{eq:app_TP_CM_standard_form} can be equivalently modeled as a single-mode Gaussian channel. We see from Eqs.~\eqref{eq:TP_parameters_MS} and~\eqref{eq:TP_parameters_OS} that $\hat{q}$ and $\hat{p}$ quadratures have unbalanced noise unless $C_v=0$. So it is necessary to introduce different scaling factors $\kappa_q$ and $\kappa_p$ in the corrective displacement in the teleportation. From Ref.~\cite{wu2021deterministic}, the Wigner function of the output state $\hat{\rho}_{\rm{out}}$ from microwave-to-optical transduction is given by
\begin{align}
    &W(\bx_B) = \int d^2 \bx d^2 \mathbf{\tilde{x}} \; W^{AB}\left[\bZ_2(\bx + \tilde{\bx}),\bx_B+\bm{\kappa}  \mathbf{\tilde{x}} \right]W^{\rm {in}}(\bx) \nonumber \\
    & \propto \int d^2 \bx  W^{\rm {in}}(\bx) \times \nonumber
    \\
    &\quad \exp{-\frac{(x_{Bq}-x_q 
    \kappa_q)^2}{u_q \kappa_q^2-2v_q \kappa_q+w_q}-\frac{(x_{Bp}-x_p 
    \kappa_p)^2}{u_p \kappa_p^2-2v_p \kappa_p+w_p}},
    \label{eq:teleportation-channel-Wigner-function}
\end{align}
where $\bm{\kappa} = \text{Diag}(\kappa_q, \kappa_p)$ and $\bZ_2=\text{Diag}(1,-1)$.  When $\kappa<1$, the channel described by Eq.~\eqref{eq:teleportation-channel-Wigner-function} has the same Wigner function transform as a Gaussian channel with two single-mode squeezing concatenated by a beam splitter with transmissivity $\kappa^2 =\kappa_q \kappa_p$, as shown in Fig.~\ref{fig:channel1} (a). The symplectic transform $S = K.\text{Sq}(r_1,r_2)$, where
\begin{align}
    & \text{Sq}(r_1,r_2)=\text{Diag}(e^{r_1},e^{-r_1},e^{r_2},e^{-r_2}), \\
    & K = \begin{pmatrix}
    \kappa \mathbf{I}_2 &\sqrt{1-\kappa^2} \mathbf{I}_2 \\
    -\sqrt{1-\kappa^2} \mathbf{I}_2 & \kappa \mathbf{I}_2
    \end{pmatrix}.
\end{align}

We provide the detailed analyses in the following.
The initial state $\hat{\rho}$ at port $1$ is independent with the state at port $2$, which is in a thermal state $\hat{\rho}_{\rm{th}}$ with mean photon number $\Bar{n}_{\rm{th}}$. The joint Wigner function
\begin{align}
W(\bx_{12}) &= W^{\rm{in}}(\bx_1) W^{\rm{th}}(\bx_2) \nonumber \\
   & \propto W^{\rm{in}}(\bx_1) \exp{-\frac{\bx_2^2}{(2\Bar{n}_{\rm{th}}+1)}}
\end{align}
They are squeezed independently by $\text{Sq}(r_1,r_2)$ and mixed by a beam splitter $K$, leading to the Wigner function
\begin{align}
&W^\prime(\bx_B,\bx_3) = W(S^{-1}\bx_{12})\nonumber\\
    & \propto W^{\rm{in}}\left(e^{-r_1}(-x_{3q}\sqrt{1-\kappa^2}+x_{Bq} \kappa),e^{r_1}(x_{3p}\sqrt{1-\kappa^2}+x_{Bp} \kappa)\right) \times \nonumber \\
    &\exp{-\frac{(x_{Bq}\sqrt{1-\kappa^2}+x_{3q} \kappa)^2}{e^{2r_2}(2\Bar{n}_{\rm{th}}+1)}}
    \exp{-\frac{(x_{Bp}\sqrt{1-\kappa^2}+x_{3p} \kappa )^2}{e^{-2r_2}(2\Bar{n}_{\rm{th}}+1)}}.
\end{align}
We get the final output $\calE (\hat{\rho})$ by tracing out the environment at port $3$. 
\begin{widetext}
    \begin{align}
    W(\bx_B) &\propto \int d^2 \bx_3 W^\prime(\bx_B,\bx_3) \nonumber \\
    &\propto \int d^2 \bx  W^{\rm {in}}(\bx) \exp{-\frac{(x_{Bq}-x_q 
    e^{r_1} \kappa )^2}{e^{2r_2}(1-\kappa^2)(2\Bar{n}_{\rm{th}}+1)}-\frac{(x_{Bp}-x_p 
    e^{-r_1} \kappa )^2}{e^{-2r_2}(1-\kappa^2)(2\Bar{n}_{\rm{th}}+1)}}.
    \label{eq:Wigner_function_xb}
    \end{align}
\end{widetext}
In Eq.~\eqref{eq:Wigner_function_xb}, we make use of the following substitution to change dummy integral values
\begin{align}
    & x_q=e^{-r_1}(-x_{3q}\sqrt{1-\kappa^2}+x_{Bq} \kappa),\\
    & x_p = e^{r_1}(x_{3p}\sqrt{1-\kappa^2}+x_{Bp} \kappa).
\end{align}
Comparing Eq.~\eqref{eq:Wigner_function_xb} with Eq.~\eqref{eq:teleportation-channel-Wigner-function}, we conclude that the teleportation channel is equivalent to a single-mode Gaussian channel with the following parameters
\begin{align}
    & \kappa^2=\kappa_q\kappa_p,\\
    & e^{r_1}=\sqrt{\kappa_q/\kappa_p},\\
    &(\Bar{n}_{\rm{th}}+\frac{1}{2})(1-\kappa^2)=\nonumber \\
    & \quad \frac{1}{2}\sqrt{(u_q \kappa_q^2-2v_q \kappa_q+w_q)(u_p \kappa_p^2-2v_p \kappa_p+w_p)} ,\\
    & e^{4 r_2} =\frac{u_q \kappa_q^2-2v_q \kappa_q+w_q}{u_p \kappa_p^2-2v_p \kappa_p+w_p}.
\end{align}
A similar derivation can be obtained when $\kappa\ge 1$, and 
\begin{equation}
    K = \begin{pmatrix}
     \kappa  \mathbf{I}_2 &\sqrt{\kappa^2-1} \mathbf{Z}_2 \\
    \sqrt{\kappa^2-1} \mathbf{Z}_2 &  \kappa \mathbf{I}_2
    \end{pmatrix}.
\end{equation}
The channel can be simplified to a Gaussian channel of class $B_2$ or $C$ with the assist of Gaussian operations ${U}_{T_1}$ and ${U}_{T_2}$ as shown in Fig.~\ref{fig:channel1} (b) \cite{Holevo2007}, where $T_1=\text{Sq}(-r_1)$ and  $T_2=\text{Sq}(-r_2)$ are single-mode squeezings.
The simplified channel has transmissivity and mean photon number given by
\begin{align}
& \eta_{\rm{TP}}=\kappa_q \kappa_p ,\\
  &\Bar{n}_{\rm{TP}} =
    \begin{cases}
      ({\rm{det}} \mathbf{N}_{\rm{TP}})^{\frac{1}{2}} & \text{if $\eta_{\rm{TP}}=1$};\\
      \frac{({\rm{det}} \mathbf{N}_{\rm{TP}})^{\frac{1}{2}}}{|1-\eta_{\rm{TP}}|}-\frac{1}{2} & \text{if $\eta_{\rm{TP}} \neq 1$ and $\eta_{\rm{TP}} > 0$ },
    \end{cases}       
\end{align}
where $\mathbf{N}_{\rm{TP}}$ is the noise matrix. We summarize them in four different cases
\begin{align}
    &\mathbf{N}_{\rm{TP,MS}}^{\rm m \veryshortrightarrow o} = \frac{1}{2}\begin{pmatrix}
u_q \kappa_q^2-2v_q \kappa_q+w_q & 0   \\
0 & u_p \kappa_p^2-2v_p \kappa_p+w_p
\end{pmatrix},\\
    &\mathbf{N}_{\rm{TP,MS}}^{\rm o \veryshortrightarrow m} = \frac{1}{2}\begin{pmatrix}
w_q \kappa_q^2-2v_q \kappa_q+u_q & 0   \\
0 & w_p \kappa_p^2-2v_p \kappa_p+u_p
\end{pmatrix},\\
    &\mathbf{N}_{\rm{TP,OS}}^{\rm m \veryshortrightarrow o} = \frac{1}{2}\begin{pmatrix}
u_q^\prime \kappa_q^2-2v_q^\prime \kappa_q+w_q^\prime & 0   \\
0 & u_p^\prime \kappa_p^2-2v_p^\prime \kappa_p+w_p^\prime
\end{pmatrix},\\
    &\mathbf{N}_{\rm{TP,MS}}^{\rm o \veryshortrightarrow m} = \frac{1}{2}\begin{pmatrix}
w_q^\prime \kappa_q^2-2v_q^\prime \kappa_q+u_q^\prime & 0   \\
0 & w_p^\prime \kappa_p^2-2v_p^\prime\kappa_p+u_p^\prime
\end{pmatrix}.
\end{align}
The teleportation channel reduces to the one discussed in Ref.~\cite{wu2021deterministic} when $C_v=0$. In that case, the noise of $\hat{q}$ and $\hat{p}$ are balanced. Due to the symmetry, we take $\kappa=\kappa_q=\kappa_p$ and recover the same results as Ref.~\cite{wu2021deterministic}.

\section{Fidelity calculation}
\label{app:fidelity}
Besides the fundamental quantum capacity, the fidelity between the input and output state is a more practical metric to characterize the performance of transduction. However, the fidelity depends on what state we transfer. Below, we provide some examples that are important in many quantum communication systems. Suppose that a single mode Gaussian channel has transmissivity $\kappa^2$ and noise $N=(\Bar{n}+\frac{1}{2})|\kappa^2-1|$. When transferring a quantum state with Wigner function $W^{\rm{in}}(\bx)$, the output state Wigner function $W^{\rm{out}}(\bx_B)$ is given by
\begin{equation}
    W^{\rm{out}}(\bx_B)=\int \diff^2{\bx} W^{\rm{in}}(\bx) \frac{1}{2N\pi}\exp{-\frac{(\kappa \bx -\bx_B)^2}{2N}}.
\end{equation}
The Uhlmann fidelity $
\mathcal{F}\left(\hat{\rho},\hat{\sigma}\right)=\left({\rm tr}\sqrt{\sqrt{\hat{\rho}}\hat{\sigma}\sqrt{\hat{\rho}}}\right)^2
$~\cite{uhlmann1976transition,jozsa1994fidelity}, which quantitatively describes the similarity between two quantum state $\hat{\rho}$ and $\hat{\sigma}$. In our case, we consider a pure input state and the fidelity further simplifies to
\begin{align}
    \mathcal{F}&=2\pi \int \diff^2{\bx_B} W^{\rm{out}}(\bx_B) W^{\rm{in}}(\bx_B).
\end{align}

\subsection{Coherent state}
A coherent state $\ket{\alpha}$ is determined by a complex number $\alpha = \alpha_R + i \alpha_I$, where $\alpha_R$ and $\alpha_I$ are real and imaginary component, respectively. We use the vector notation $\balpha = \sqrt{2}(\alpha_R , \alpha_I)^{\rm T}$, then its Wigner function is well-known
\begin{equation}
    W^{\rm{in}}(\bx)=\frac{1}{\pi} \exp{-(\bx-\sqrt{2}\balpha)^2},
\end{equation}
where we use the short-handed notation $\bx^2$ as the square of the norm of the vector $\bx$.
If the single-mode Gaussian channel has the noise being mixed in $N$ and transmissivity $\kappa^2$, the fidelity between the input and output state can be obtained as
\begin{align}
    \mathcal{F}&= \frac{2}{1+2N+\kappa^2}\exp{-\frac{2\balpha^2(\kappa-1)^2}{1+2N+\kappa^2}}.
\end{align}

\subsection{Cat state}
The cat states are defined as $\ket{cat_{\pm}}\equiv N_{\pm}(\ket{\alpha}\pm\ket{-\alpha})$, where $N_{\pm}=(2\pm 2 e^{-2\balpha^2})^{-1/2}$ are normalization constants. The corresponding Wigner functions and the fidelity between the input and output are derived similarly:
\begin{widetext}
\begin{align}
    &W^{\rm {in}}(\bx;\ket{cat_{\pm}})=N_{\pm}^2\frac{1}{\pi}\left[e^{-(\bx-\sqrt{2}\balpha)^2}+e^{-(\bx+\sqrt{2}\balpha)^2} \pm 2 e^{-\bx^2}\cos[2\sqrt{2}(-q\alpha_I+p\alpha_R)]\right],\\
    &\mathcal{F} = \frac{4N^4_{\pm}}{1+2N+\kappa^2}\left(e^{-\frac{2\balpha^2(1-\kappa)^2}{1+2N+\kappa^2}}+e^{-\frac{2\balpha^2(1+\kappa)^2}{1+2N+\kappa^2}} \pm 2e^{-\frac{2\balpha^2(2+2N)}{1+2N+\kappa^2}}
    \pm 2e^{-\frac{2\balpha^2(2N+2\kappa^2)}{1+2N+\kappa^2}}+e^{-\frac{2\balpha^2[4N+(1+\kappa)^2]}{1+2N+\kappa^2}}+e^{-\frac{2\balpha^2[4N+(1-\kappa)^2]}{1+2N+\kappa^2}}\right).
\end{align}
\end{widetext}

%

\end{document}